\documentclass[11pt,a4paper]{article}
\pdfoutput=1
\usepackage{jheppub}

\usepackage{color}
\usepackage{amsmath}
\usepackage{pifont}

\usepackage{bbm}
\usepackage{verbatim}   
\usepackage{subfigure}  
\usepackage{acronym}

\usepackage{amsfonts}
\usepackage{amssymb}
\usepackage{mathrsfs}
\usepackage{graphicx}
\usepackage{multirow}
 \usepackage{slashed}

\newcommand{\ie}{{ i.e.}}

\definecolor{mygreen}{RGB}{29,145,47}
\definecolor{mypurple}{RGB}{164,64,214}
\definecolor{myorange}{RGB}{199,146,32}

\newcommand{\tn}{\tilde{n}}

\newcommand{\second}{{\, {\rm sec}}}
\newcommand{\fermi}{{\, {\rm fm}}}
\newcommand{\barn}{{\, {\rm barn}}}
\newcommand{\pb}{{\, {\rm pb}}}

\newcommand{\Gyr}{{\, {\rm Gyr}}}

\newcommand{\cm}{{\, {\rm cm}}}

\newcommand{\keV}{{\, {\rm keV}}}
\newcommand{\MeV}{{\, {\rm MeV}}}
\newcommand{\GeV}{{\, {\rm GeV}}}
\newcommand{\TeV}{{\, {\rm TeV}}}

\def\beq{\begin{equation}}
\def\eeq{\end{equation}}
\def\bea{\begin{eqnarray}}
\def\eea{\end{eqnarray}}
\def\bitem{\begin{itemize}}
\def\eitem{\end{itemize}}
\newcommand{\bec}{\begin{center}}
\newcommand{\eec}{\end{center}}
\newcommand{\ba}{\begin{array}}
\newcommand{\ea}{\end{array}}

\def\bar#1{\overline{#1}}

\def\inv{^{\raise.15ex\hbox{${\scriptscriptstyle -}$}\kern-.05em 1}}
\def\lbar{{\lower.35ex\hbox{$\mathchar'26$}\mkern-10mu\lambda}} 

\let\<=\langle
\let\>=\rangle

\let\+=\uparrow

\begin{document}

\title{Big Bang Synthesis of Nuclear Dark Matter} 
\author[a,b]{Edward Hardy,}
\emailAdd{ehardy@ictp.it}
\author[a]{Robert Lasenby,}
\emailAdd{robert.lasenby@physics.ox.ac.uk}
\author[a,c]{John March-Russell,}
\emailAdd{jmr@thphys.ox.ac.uk}
\author[d]{Stephen M. West}
\emailAdd{stephen.west@rhul.ac.uk}
\affiliation[a]{Rudolf Peierls Centre for Theoretical Physics, University of Oxford, 1 Keble Road,\\
Oxford, OX1 3NP, UK}
\affiliation[b]{Abdus Salam International Centre for Theoretical Physics,
Strada Costiera 11, 34151, Trieste, Italy}
\affiliation[c]{Stanford Institute for Theoretical Physics, Department of Physics, Stanford University,\\
Stanford, CA 94305, USA}
\affiliation[d]{Physics Department, Royal Holloway, University of London,
Egham, Surrey, TW20 0EX, UK}

\abstract{
We investigate the physics of dark matter models featuring composite
bound states carrying a large conserved dark ``nucleon'' number.
The properties of sufficiently large dark nuclei may obey simple scaling laws,
and we find that this scaling can determine the number distribution
of nuclei resulting from Big Bang Dark Nucleosynthesis.
For plausible models of asymmetric dark matter, dark nuclei of large nucleon number,
e.g.\ $\gtrsim 10^8$, may be synthesised, with the number distribution
taking one of two characteristic forms. 
If small-nucleon-number fusions are sufficiently fast, the
distribution of dark nuclei takes on a logarithmically-peaked,
universal form, independent of many details of the initial conditions
and small-number interactions.
In the case of a substantial bottleneck to nucleosynthesis for small dark
nuclei, we find the surprising result that even larger nuclei,
with size $\gg 10^8$, are often finally synthesised, again with a simple
number distribution.  We briefly discuss the constraints arising from
the novel dark sector energetics, and the extended set of (often
parametrically light) dark sector states that can occur in complete models
of nuclear dark matter.  The physics of the coherent enhancement of direct detection signals,
the nature of the accompanying dark-sector form factors, and the possible
modifications to astrophysical processes are discussed
in detail in a companion paper.
}

\maketitle


\section{Introduction}
\label{sec:intro}

Most models of dark matter (DM) assume that it is made up of point-like
particles, or at least can be treated as such in its interactions.
However, this is not necessarily the case.  In the Standard Model (SM)
most of the SM-sector mass in the universe is in the form of composite states---atoms, 
containing nuclei, made up of nucleons, which are themselves made up of
quarks and gluons.  Particularly noteworthy is the remarkable
richness of SM-sector nuclear and atomic physics resulting from just a few basic conservation laws, most importantly
baryon-number and charge conservation, and a few relevant interactions, dominantly the short-range
strong nuclear binding force and long-range electromagnetic interaction.
The compositeness and variety of the states that result, atoms and nuclei, plays a vital role in many important physical
processes in the history of the universe, for example, in galaxy formation and
all of stellar physics.  Given the ease with which bound states can arise, it is clearly important
to consider the possibility that DM may exist in the form of composite states too. 

In this work, we will investigate a particular (and as we will
argue, very attractive) class of composite DM models---those
featuring cosmologically stable ``dark nuclei'' (DN) of large
dark ``nucleon'' number (DNN). In particular we focus on the case
where, similar to SM nuclei, there is a relatively short-range
strong ``nuclear'' binding force with a hard core repulsion,
and there is an approximately-conserved quantum number, DNN,
analogous to baryon number. Unlike the SM where both protons and
neutrons are relevant, for reasons of minimality and simplicity
we will take there to be just one kind of stable dark nucleon
(or more generally, the differences between different nucleon types
to be unimportant).
Importantly, we will in addition assume that {\em in the dark sector the
analogue of the  long-range electromagnetic interaction between protons
is absent}, either because there is no massless ``dark photon''
or because the stable dark nucleon is uncharged.\footnote{Models
of composite DM previously considered in the literature include
``WIMPonium"~\cite{Pospelov2008,MarchRussell2008,Shepherd2009},
``dark atoms" \cite{Kaplan:2009de}, and most similarly to our
work, ``Q-balls"~\cite{Frieman1988, Kusenko1997} (non-topological
solitons of scalar fields). In addition the
DM candidates in technicolor-like theories are often composite
states~\cite{Nussinov:1985xr, Chivukula:1989qb}, though these typically
have small ``nucleon" number, and most importantly have constituents with
SM charges, unlike the cases we consider in this work.}

With these simple assumptions there exist a very broad range of stable
DN states of varying DNN, and with a binding energy per dark nucleon
that {\it saturates} at a finite value. This is similar to the SM where
there are stable nuclei at multiple baryon numbers with approximately
constant binding energy per nucleon, $\sim 8 \MeV$, over a range of
nuclear sizes. Unlike the SM, however, there is no longer a Coulomb
repulsion term, so that in the dark sector the binding energy per
dark nucleon truly does saturate, at least until gravitational effects become
important, and never turns over, unlike the Fe maximum in the SM, so
there  are {\em stable DN up to very large DNN}. In terms of a simple
liquid-drop like model, analogous to the semi-empirical mass formula
in the SM, we take the binding energy per dark nucleon to behave
asymptotically as
\begin{equation}
\frac{B_k}{ k }= \alpha  - \frac{\beta}{k^{1/3}} ,
\label{eq:semf}
\end{equation}
where $k\gg 1$ is the DNN, the volume energy term is $\alpha$, and the
surface tension term $\beta$. Roughly this set of assumptions was
considered in~\cite{Krnjaic2014}, which investigated the possibility that
they could result in the early-universe formation of large DN.~\footnote{Essentially
this holds for SM nuclei if Coulomb repulsion is absent (see \cite{Krnjaic2014} for such a model), and also for many 
Q-ball models~\cite{Frieman1988,Frieman1989,Kusenko1997}.  In the absence of short-range
hard-core repulsion, one obtains, in the fermionic case, states such
as those investigated in~\cite{Wise2014,Wise2014ii}.  These are supported by degeneracy
pressure until their constituents become relativistic,
at which point the form of the attractive interactions becomes important---they 
will generally not display the kind of scaling we consider in this paper.}

While the small bound states of such a model may have
highly variable properties,\footnote{See~\cite{Detmold2014,Detmold2014ii}
for a detailed investigation of a particular QCD-like model
focussing on the properties of the small bound states.} analogous
to the special properties of D or $^4$He, once the DN become large enough,
their properties, beyond just their binding energy, eq.(\ref{eq:semf}), can
obey simple scaling laws.  Most importantly certain interaction cross
sections scale geometrically in the large DNN limit, at least when
suitably averaged over a range of DNN to eliminate ``magic number'' and resonant
effects which are of subdominant importance to the physics we study. 

The main focus of this paper is the regime in which such scaling 
properties can determine the final number distribution of DM;
namely, when larger states are built up via the 
agglomeration of smaller ones at low temperatures, as per SM Big Bang Nucleosynthesis (BBN).
Like the BBN case, the simplest version of this process
applies when the DM is {\it asymmetric} \cite{Nussinov:1985xr,Gelmini:1986zz,Chivukula:1989qb,Barr:1990ca,Kaplan:1991ah,
Thomas:1995ze,Hooper:2004dc,Kitano:2004sv,Agashe:2004bm,Cosme:2005sb,Farrar:2005zd,Suematsu:2005kp,Tytgat:2006wy,Banks:2006xr,
Khlopov:2008ty,Kitano:2008tk,Kaplan:2009ag,Kohri:2009yn,Kribs:2009fy,Cohen:2009fz,An:2009vq,Khlopov:2010pq,Cohen:2010kn,Shelton:2010ta,
Davoudiasl:2010am,Haba:2010bm,Buckley:2010ui,Hall:2010jx,Dutta:2010va,Falkowski:2011xh,Frandsen:2011kt,Graesser:2011wi,Buckley:2011kk,
Iminniyaz:2011yp,MarchRussell:2011fi,MarchRussell2012,Hardy:2014dea,Unwin:2014poa}, \ie, with a particle--anti-particle
asymmetry that determines the final DM abundance, similarly to the baryon
asymmetry in the SM sector (though the symmetric case is also interesting).
The small-DN part of such a scenario was investigated in~\cite{Krnjaic2014}.
In the following we assume, unless specifically stated otherwise, the DM to be asymmetric
so the DNN will take positive integer values, and we denote the DN of various DNN as 1-DN, 2-DN, etc.

In the main body of this paper, Sections~\ref{sec:dn} and~\ref{sec:agg}, we will show that the
dark-sector analogue of SM BBN---which we call
``Big-Bang Dark Nucleosynthesis'' (BBDN)---can produce DN with DNN
up to or even exceeding $10^8$, resulting in a very wide variety of
striking changes to traditional DM phenomenology.

Specifically, we find the number distributions of DN resulting from BBDN
satisfy remarkably simple forms. For example, for asymmetric DM with
plausible underlying parameter values, DN with DNN $\gtrsim
10^8$ can be synthesised, with the number distribution taking one of two
characteristic scaling forms. If there is no substantial bottleneck
of BBDN at small DNN the distribution of DN sizes takes on a
peaked, in log space,
universal non-power-like functional form, independent of many details
of the initial distribution and interactions. This solution acts as an
attractor solution and we study how the distribution function of DN
approaches this universal scaling form. On the other hand, in the case
of a substantial bottleneck of BBDN at small DNN we find the surprising
result that DN of even larger DNN, $\gg 10^8$, are often synthesised,
again with a simple form of the number distribution. Such behaviours are
studied both via numerical solutions of the relevant equations, and from
physically motivated approximate analytical solutions.

\vspace{0.1in}
Such states can give rise to a variety of
interesting possibilities, including:
\begin{itemize}

\item For reasonable parameter values effectively very heavy ($\ge
10^8 \GeV$) DM can be produced by BBDN in the form of large DN (and
with one of two possible, essentially model independent
distribution functions). Such heavy DM is in contrast to the usual
unitarity limit for point-like DM in the case of thermal freeze-out
production \cite{Griest:1989wd},\footnote{Though thermal {\em freeze-in}
production of DM can produce superheavy elementary DM~\cite{Hall:2009bx} with an
energy density yield independent of the superheavy particle mass.}
and is not usually expected in asymmetric DM models, which seek to link the
DM and visible sector abundances.

\item Coherent enhancement of interactions: Processes that
do not probe the individual constituents will have amplitude
going as the DNN $k$, so scattering cross sections, in for example direct detection
experiments, can in principle be enhanced by $k^2\gtrsim 10^{16}$ compared to the case of a single
DM nucleon.  Taking account of the reduction in the number density of such states, one still
finds an effective increase in direct detection interaction rates
scaling as $\sim k$, so effectively {\it reducing} expected collider signals
by $\sim 1/k$ for a given direct detection rate compared to the standard
point like case.
The physics of the coherent enhancement of direct detection signals, with
the accompanying possibility of novel form factors from the dark sector,
is discussed in detail in a companion paper \cite{NDMdirect}.

\item Inelastic processes at both ``high energy'', of order the DN
binding energy differences, and parametrically lower energy, from
long-wavelength collective excitations. Examples of the high energy
processes are ones that move between states of different DNNs---fusions and fissions,
but there is also the possibility of the extended structure of states
leading to a spectrum of excitations, as exists in SM nuclei and atoms
\cite{NDMdirect}.

\item States with large spin: for large composite states,
there is the possibility of interactions aligning the spins
of many of the constituent states, leading to DN with nuclear spin $\sim k$. 
This is of potential interest in, for example, interactions with SM nuclei
in direct detection experiments, and capture in astrophysical objects.
\end{itemize}
Since our focus is the physics of BBDN, many of the above possibilities
are either only briefly touched upon in this work, or not treated at
all, and we leave their detailed study to a series of companion papers.

Finally, before turning to our analysis of DN and BBDN we emphasise that while there are many
specific models which can realise this kind of scenario, we will focus on regimes in which
the behaviour can be broadly model-independent in nature.


\section{Basics of dark nucleosynthesis}
\label{sec:dn}

While present-day DM may be composed of large bound states,
this is generically not the case in the hot early universe.
At large temperatures $T$, the entropy term in the free energy
dominates and the chemical equilibrium distribution
has almost all DM in small-number states.
For small $T$, compared to the binding energies, the energy term
dominates, and chemical equilibrium favours large bound states.

As we discuss in detail in later sections, starting from the situation at high temperatures,
large DN may be assembled by an aggregation process where fusions dominate dissociations and
fissions. Specifically, if interactions are not so weak
as to be frozen out by the time the equilibrium shifts to favour larger
states, then larger states will be built up until fusion reactions
freeze out due to a combination of falling velocity and a falling
number density from both Hubble expansion and the formation of the large DN themselves. 

Since, for the DM masses we consider, the DM is dilute, if we are in kinetic equilibrium
then the transition from being kept in equilibrium by dissociations, to fusion
reactions dominating, generally happens fast enough to be only a small perturbation
to the subsequent fusion process (this technical point is discussed in detail in
Appendix~\ref{app:chemeq}). If thermalising interactions 
are not sufficiently fast to reduce the energy of de-excitation
products before they hit another DN, then these may cause dissociations,
leading to very different behaviour from the fusions-only approximation
(c.f.\ SM recombination).\footnote{\label{fn:dex}In principle,
another exception to the statement that, late on, only fusions are important 
is when the excited states produced by the fusion processes de-excite
by the emission of nucleon constituents, as occurs
in the SM~\cite{Bertulani2007} (many Q-ball models also de-excite by losing
charge, e.g.~\cite{Multamaki:2000qb}).
In general these emitted small-$k$ DN are either
quickly re-absorbed by larger DN and do not significantly alter
the dynamics of the aggregation process, or they act approximately
as an external bath with which the larger DN scatter. In the latter case,
as long as enough of the small-small fusion processes can occur without involving nucleon emission,
the mass fraction of small DN will be sub-dominant,
though they may dominate the number density.}
We will not consider this regime in the current
paper, assuming instead that the de-excitation products decay or thermalise
on fast enough timescales.\footnote{For example, if there is a bath
of states which a given de-excitation product interacts with more frequently
than it does with the DN, then the de-excitation products
will thermalise with that bath quickly
(cf.\ the SM electron bath during SM BBN).
Alternatively, the de-excitation products may decay quickly to lighter
hidden sector states (in particular, as discussed in Section~\ref{sec:lightdark},
if the de-excitation products are lighter than $\sim 100 \MeV$ then they generally
need to interact with lighter hidden sector states to reduce their
cosmological density).}


\subsection{Freeze-out of fusions}
\label{sec:fusionfo}

Given that fusions dominate at low $T$, we can obtain an estimate
of the maximum size of DN built up by the aggregation process
by investigating when fusion reactions freeze out.

First, let us suppose that the last fusions to freeze out
are those between large DN, and also that we end up with
a `peaked' mass distribution in which almost all of the mass is in $\sim A$-DN.
In this case, the rate of fusions for a single DN, per Hubble time,
is $\Gamma/H \sim \langle \sigma v \rangle n_A / H$,
where $\langle \sigma v \rangle$ is the thermally-averaged fusion
cross section, and $n_A$ is the number density of $A$-DN.
Since we have DNN conservation, $n_A = n_0/A$, where
$n_0$ is the total DNN density.
If the DN are non-relativistic, as required to be aggregating,
and in thermal equilibrium (e.g., with an external bath of light dark-sector particles as
we discuss later), then the DN velocity $v_A \simeq v_1 A^{-1/2}$. 
For large $A$, saturation of the dark sector nuclear forces implies the internal mass-energy
density, $\rho_b$, of DN matter is roughly constant with size, and that fusion cross
sections scale approximately geometrically, $\sigma \simeq \sigma_1 A^{2/3}$, and so
\begin{equation}
\frac{\Gamma}{H} 
\sim \frac{\sigma_1 v_1 n_0}{H} A^{2/3} A^{-1/2} A^{-1}
= \frac{\sigma_1 v_1 n_0}{H} A^{-5/6} \,.
\label{eq:foscaling}
\end{equation}
In general, the temperature, $T_b$, of the DN bath can differ from the temperature, $T$, which sets the 
Hubble expansion rate in the radiation dominated era, and which we assume to be dominantly set by the
more numerous SM sector degrees of freedom, as the dark and SM sectors may be essentially decoupled
from each other.  

With this in mind, and using $\sigma_1 \sim 4 \pi R_1^2$,
$v_1^2 \sim T_b / m_1$, and $4\pi R_1^3 \rho_b/3 \sim m_1$,
we then find
\begin{equation}
\frac{\sigma_1 v_1 n_0}{H} \sim
2 \times 10^7 \, \left(\frac{1 \GeV \fermi^{-3}}{\rho_b}\right)^{2/3}
\left(\frac{g_\star(T)}{10.75}\right)^{1/2}
\left(\frac{m_1}{1 \GeV}\right)^{-5/6}
\left(\frac{T}{1 \MeV}\right)^{3/2} 
\left(\frac{T_b}{T}\right)^{1/2} ,
\label{eq:fon1}
\end{equation}
where the parameters are normalised to SM values both for comparison and
because such a parameter region is naturally motivated by asymmetric
dark matter (ADM) models.
Thus, in this scenario, if dissociation stops being important
around $T = 1 \MeV$, then, from eq.(\ref{eq:foscaling}), the largest mass that could have been built up
is $(2 \times 10^7)^{6/5} m_1 \sim 5 \times 10^8 \GeV$,
corresponding to a radius of $\sim 480\fermi$ for the fiducial parameter values.\footnote{Note that this does
not depend on the dark nucleon mass $m_1$. If some scaling other than $m_k\propto k$ between mass and
DNN held, e.g.\ as in the case of some Q-ball models~\cite{Frieman1989},
then eq.(\ref{eq:foscaling}) would still apply, but with $A$ replaced by the ratio of the final
mass to the mass for which $\sigma_1 v_1 n_0$ was evaluated.
Such models, in which binding energies were not a small fraction of the total mass,
have quite different properties.}
Beyond this size, the number density and velocity are too low for interactions
to occur.  We will see that, for reasonable parameters, this bound
may be attained.

Note that, if we scale all
of our dimensional parameters to increase mass scales by a factor
$\lambda$ (i.e.\ $T \mapsto \lambda T$, $\rho_b \mapsto \lambda^4 \rho_b$, etc.),
then the freeze-out DN mass scales as $m_{{\rm fo}} \mapsto \lambda^{-7/5} m_{{\rm fo}}$. 
Changing the mass scales of the constituents, e.g.\ by changing the confinement
scale in a strongly coupled theory, may be expected to have roughly this effect.
Thus, {\em decreasing} the mass scale of our constituents will tend to
{\em increase} the masses that could be built up, mainly since larger
geometric cross sections are available.
Going the other way, for $m_1 \gtrsim 100 \TeV$ (and corresponding 
scalings of the other parameters), we have $\sigma_1 v_1 n_0 / H < 1$,
so there is no build-up in this regime, corresponding to 
the usual unitarity bound for DM self-annihilations.

Alternatively, the last fusions to freeze out may be those
between `small' $+$ `large' DN.\footnote{The precise meaning of `small'
will depend on the aggregation dynamics, and as it becomes comparable
to the eventual sizes obtained, this scenario will approach the large $+$ large
case discussed above.}
Compared to large + large fusions,
while the possible number density of small DN is larger,
the number of separate fusion events onto a given DN needed for
the same increase in size is larger by the same factor.
However, since in thermal equilibrium the velocities
of small DN are larger, the overall rate of size
increase is {\em enhanced} by that factor.
To qualitatively understand this scenario, suppose that an order one proportion
of the dark nucleons are in small $k$-DN, and the rest in large $A$-DN.
Then, the rate at which an $A$-DN increases in size by $A$ is
given by
\begin{equation}
\Gamma \sim \langle \sigma v \rangle n_k \frac{k}{A}
\sim \delta \frac{1}{4} \sigma_1 v_1 n_0 k^{-1/2} A^{2/3} A^{-1} ~ ,
\label{eq:fobneck}
\end{equation}
where we used $\langle \sigma v\rangle_{A + k} = \frac{1}{4} \delta \sigma_1 v_1 k^{-1/2} A^{2/3}$,
with $\delta$ parameterising a possible suppression of small-large cross-sections
from the geometric value, e.g.\  from quantum reflection effects, as per
SM nucleon-nucleus interactions.  So, the maximum attainable DNN of the DN is
\begin{equation}
A_{\rm fo} \sim 7 \times 10^{19} \left( \frac{\delta}{\sqrt{k}} \,
\frac{\sigma_1 v_1 n_0/H}{2 \times 10^7}\right)^3 ,
\end{equation}
corresponding to $m \sim 7 \times 10^{19} \GeV$ (and
scaling as $m_{{\rm fo}} \mapsto \lambda^{-5} m_{{\rm fo}}$) and
a radius of $3 \times 10^6 \fermi$ for our fiducial parameter choices.

Similarly, the rate of fusions onto an $A$-DN for a given
$k$-DN is $\Gamma \sim \langle \sigma v \rangle n_A 
\sim \frac{n_A}{n_k} \langle \sigma v \rangle n_k$,
so the largest DN that can absorb the entire population
of small DN also have approximately this same maximum size.
Thus, as long as $\delta$ is not too small,
and a large enough population of small DN exists
for long enough, there is the possibility
of producing much larger DN via this route, 
than via an aggregation process in which the DN at any given
time are of approximately the same size.

Obtaining significant quantities of DN this large requires that a
population of small DN remains around until the end of the aggregation process,
i.e.\ that small $+$ small fusions are slow. If this is the case, but
small $+$ large fusion cross sections are roughly geometrical, and we
produce a number density of `seed' large DN $\sim n_0/A_{\rm fo}$, then,
as studied in Section~\ref{sec:bn}, we will end up with most of
the DM mass in DN of size $\sim A_{\rm fo}$. If we produce a larger seed density, the
maximum size will be reduced proportionally, up to a cross-over point at
which this size is lower than the freeze-out limit for large $+$ large
fusions, at which point we enter that regime. If a smaller
seed density is produced, then most of the DN never gets through the `bottleneck',
and remains small, with some sub-dominant population of large DN up to
$m_{\rm fo}$. Section~\ref{sec:bn} investigates these regimes in detail.


\subsection{Bottlenecks, and comparison to BBN}

Bottlenecks to nucleosynthesis, in the form of
`anomalously slow' small + small fusion rates for certain
channels, are important in Standard Model BBN.
There, only small nuclei are synthesised, with almost all
of the nucleons ending up in H and $^4$He, only small 
amounts in the other $A\leq 7$ nuclei, and entirely negligible amounts
beyond this. The $B_D \sim 2 \MeV$ binding energy of
D means that, assuming only $A=1,2$ states are occupied,
$n_D/n_p \sim \eta (T/m_p)^{3/2} e^{B_D/T}$,
and since the baryon to photon ratio $\eta \simeq 6 \times 10^{-10}$,
this only becomes $\simeq 1$ at $T \simeq 0.06 \MeV$.
However, slightly before this temperature,
the D number density becomes large enough that $2+1$ and other processes
start occurring, and in fact their rates are $\gtrsim 10^4$
times the Hubble rate, as expected from a calculation along the lines of 
eq.(\ref{eq:fon1}).

The real bottleneck preventing the synthesis of large
nuclei in the SM is the large binding energy per nucleon of $^4$He compared to
subsequent small nuclei.  It is not until $^{12}$C that the binding energy per nucleon
exceeds $^4$He.  In quasi-equilibrium among small-number nuclei, this large binding
energy means that $^4$He dominates the mass fraction along
with H, their ratio being set by the abundance of neutrons. The small number densities
of $A=7$ nuclei produced are nowhere near sufficient to make the rate of 
production of energetically favoured $A \ge 12$ nuclei comparable
to the Hubble rate, and so we freeze out with
quasi-equilibrium abundances (for a textbook discussion see e.g.~\cite{Mukhanov2003}).
Note that this kind of bottleneck, involving a wrong-sign
binding energy difference rather than merely a small right-sign
difference as in the D example, gets worse rather than better
with decreasing temperature.\footnote{In the D case,
since $n_D$ increases exponentially
with falling temperature, the number of $A \ge 3$ states produced
beforehand is insufficient to bypass the bottleneck.
For this reason, it is somewhat difficult for the bypass process
to be dominant for right-sign binding energy differences.}
In summary, SM BBN provides an example of a nucleosynthesis
process where the presence of large binding energy differences among
small-number states creates a bottleneck. In fact, there are two
bottlenecks, the first of which, the D bottleneck, we get through well
before fusions have frozen out, but not the second, post-$^4$He.

A bottleneck in BBDN may similarly have the result of
preventing significant quantities of large DN from being formed.
However, there is also the possibility, as discussed in the previous section,
of a suitable bottleneck leading to the build-up of
even larger DN than would otherwise have been produced,
with a qualitatively different distribution of DN sizes.
Section~\ref{sec:bn} explores this scenario in detail.


\section{Aggregation process}
\label{sec:agg}

Recapping, the cross-over from the high-$T$ regime to that in which
dissociations are unimportant occurs in a sufficiently short time that only DN much smaller
than the freeze-out size are able to be built up in appreciable quantities during this period.
After this, there will be a period of effectively fusion-only aggregation
before we reach one of the limits discussed in Section~\ref{sec:fusionfo}.
This section investigates the details of this aggregation process.

Evolving with fusions only, the
Boltzmann equation for the $k$-DN number density, $n_k(t)$, is
\begin{equation}
\frac{d n_k(t)}{d t} + 3 H(t) n_k(t) = -\sum_{j = 1}^\infty \langle \sigma v\rangle_{j,k}n_j(t) n_k(t)
+ \frac{1}{2}\sum_{i + j = k}\langle \sigma v \rangle_{i,j} n_i (t) n_j(t)\,,
\end{equation}
where $\langle \sigma v\rangle_{i,j}$ is the velocity-averaged
cross section for $i + j \rightarrow (i+j)$.\footnote{When de-excitation
from some fusion events is via nucleon emission,
then, as mentioned in footnote~\ref{fn:dex}, the behaviour
may still approximately follow the fusions-only aggregation solution.
If large + large fusions are the dominant process,
as in the `scaling' regime discussed in Section~\ref{sec:scaling},
then the behaviour of small DN is not crucial.
If small + large fusions are most important, as in the
`bottlenecked' regime~\ref{sec:bn}, and these dominantly
de-excite via nucleon emission, this will just reduce their
net forward rate by some factor, and, in simple cases,
leaves the qualitative behaviour otherwise intact.}

This equation combines together the dilution of number densities
due to Hubble expansion, the change in cross sections due to 
the decrease in DN velocities, and the change in the relative concentrations
of $k$-DN. We can separate these processes out by moving to different
variables. Writing the velocity-averaged cross sections as
$\langle \sigma v \rangle_{i,j} = \sigma_1 v_1 K_{i,j}$,
where the $K_{i,j}$ encode the relative rates of different fusion processes,
the Boltzmann equation in terms of the dimensionless yields $Y_k \equiv n_k/s$,
where $s(t)$ is the entropy density, is
\begin{equation}
\frac{d Y_k(t)}{dt} = \sigma_1 v_1(t) s(t) \left(
- Y_k(t) \sum_j K_{j,k}(t) Y_j(t) + \frac{1}{2} \sum_{i+j=k} K_{i,j}(t) Y_i(t) Y_j(t)\right) \,.
\end{equation}
Here we assume that entropy in conserved throughout the aggregation
process, so $s \propto a^{-3}$. This system of aggregation equations can be solved
in terms of relative concentrations, $y_k \equiv Y_k/Y_0$ (with $Y_0 = n_0/s$ the yield of dark
\emph{nucleons} irrespective of how they are distributed among DN of different sizes) and by
defining a new, dimensionless time $w$:
\begin{equation}
\frac{d y_k}{d w} = - y_k \sum_j K_{j,k} y_j + \frac{1}{2} \sum_{i+j=k} K_{i,j} y_i y_j \,,
\label{eq:agg}
\end{equation}
where
\begin{equation}
\frac{dw}{dt} = Y_0 \sigma_1 v_1(t) s(t) = n_0(t) \sigma_1 v_1(t) \, .
\label{eq:wt}
\end{equation}
Note that $Y_0$ is constant in time (assuming no entropy injection) by nucleon number conservation, and 
the solution is normalised such that $\sum_k k \, y_k = 1$ throughout.

In words, the function $w(t)$ describes how quickly the number distribution of
DN changes, whereas the set of distributions moved through is determined
by the form of the `aggregation kernel' $K_{i,j}$ and the
corresponding solution to the aggregation system eq.(\ref{eq:agg}).


\subsection{Scaling regime}
\label{sec:scaling}

The system of aggregation equations, eq.(\ref{eq:agg}) for all $k$, is simplest to solve when the
$K_{i,j}$ do not depend on $w$, i.e.\ when we can absorb all of the time/temperature
dependence into the single quantity $v_1(t)$. The simplest case
in which this is true is when the DN are in kinetic equilibrium with each
other, at some temperature $T_b(t)$. $k$-DN then have mean-square
velocities $\langle v^2 \rangle = T_b/m_k = k^{-1} (T_b/m_1)$.
If fusion cross-sections scale approximately geometrically
for large DN, then a kernel of the form
\begin{equation}
K_{i,j} = (i^{2/3} + j^{2/3})\left(\frac{1}{i^{1/2}} + \frac{1}{j^{1/2}} \right) \,,
\label{eq:ker1}
\end{equation}
is a good approximation for large DN.\footnote{This specific form
matches cases investigated in the mathematical
literature~\cite{Leyvraz2006}.} In this expression we have, for
simplicity, replaced the relative velocity, which is non-relativistic in
the fusion regime, by an averaged velocity.

\begin{figure}
\begin{center}
\includegraphics[width=0.49\textwidth]{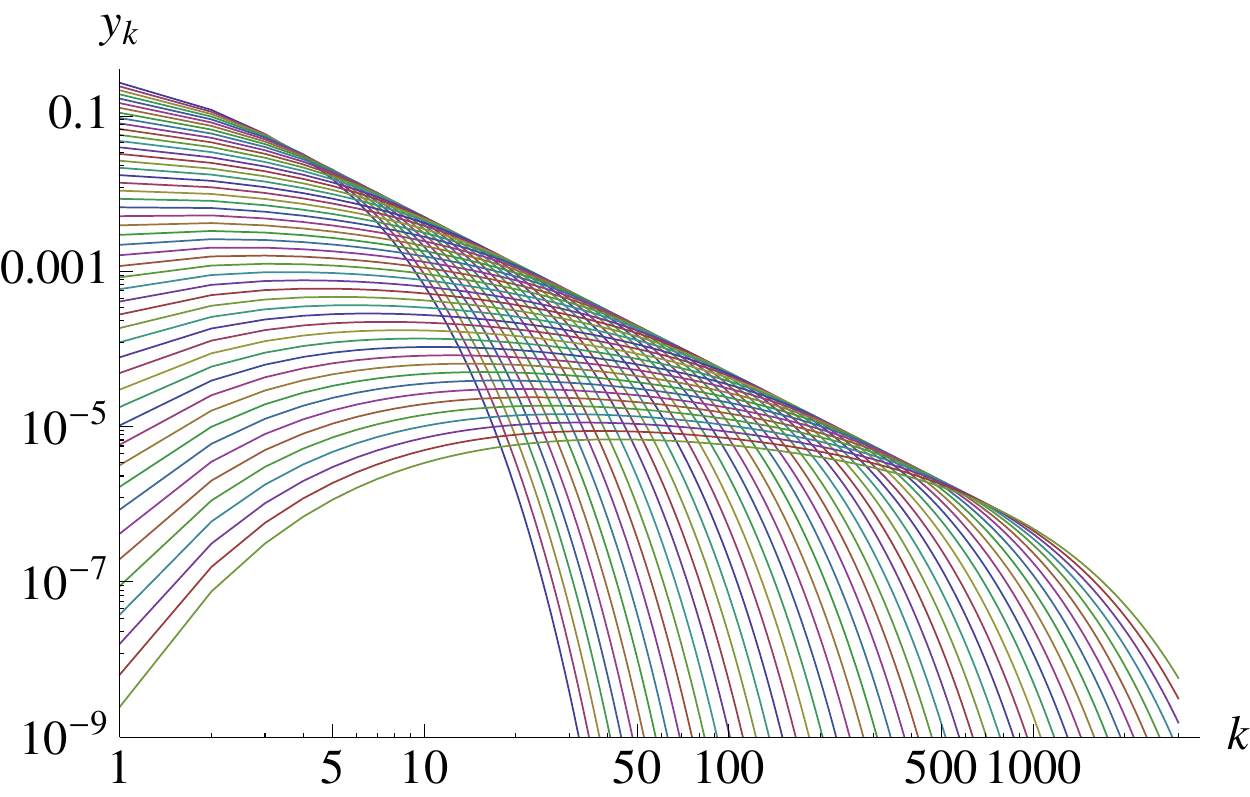}
\includegraphics[width=0.49\textwidth]{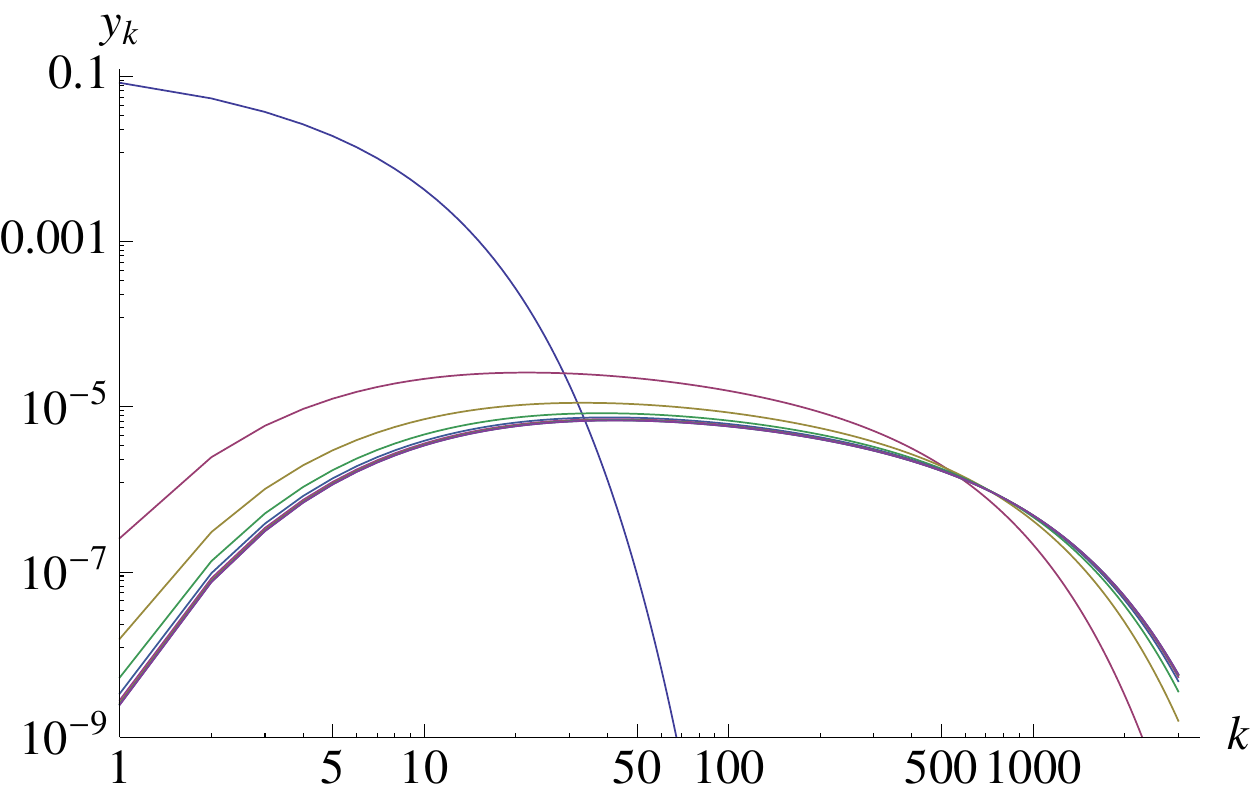}
\caption{\emph{Left:} 
Number distributions obtained by solving the system of equations
determining aggregation dynamics, eq.(\ref{eq:agg}) for all $k$,
from single-nucleon initial conditions, $y_k(0) = \delta_{k1}$, at equally-spaced
values in $\log w$ up to a maximum of $w = 75$.  The first curve is a Kronecker delta
$y_1(0) = 1$, the second curve is the one with smallest intercept on the $k$ axis, etc.
Note that distribution of DN sizes very quickly broadens in $w$-time.
\emph{Right:} The same number distributions at half $e$-folding time
intervals, assuming DN temperature falling with scale factor and $w(t)$ evolving
as per eq.(\ref{eq:wTrel}), with $w_{\max} = 75$.  As the physical time goes to infinity
the distribution is given by the $w=w_{\max}$ distribution shown by the thick solid curve.}
\label{fig:ndist}
\end{center}
\end{figure}

Figure~\ref{fig:ndist} shows, in both $w$ and $t$ space, numerical solutions for the
aggregation dynamics eq.(\ref{eq:agg}) with this kernel, starting from $y_1(0) = 1$
initial conditions. What is immediately noticeable is that the number distributions at different
times have almost the same shape, but shifted relative
to each other corresponding to an increase in average size. This arises
because our kernel is \emph{homogeneous}, $K_{b i, b j}
= b^\lambda K_{i,j}$ with, here, $\lambda = 1/6$. Physically,
if we scale up the DNN of all of the DN by some factor, this
just corresponds to an \emph{overall} scaling of the rate of fusions, and
doesn't change the relative rates of different-number processes. 
It is known \cite{RednerBook} that, for such kernels, there is generally a `scaling solution'
such that almost any finitely-supported initial conditions eventually converge
to the form 
\begin{equation}
y_k(w) = \frac{1}{\bar{k}(w)^2} f\left( \frac{k}{\bar{k}(w)}\right) \,,
\end{equation}
where $\bar{k}(w) \propto w^{1/(1-\lambda)}$ is the `characteristic size' of DN at time $w$,
$f$ is the kernel-dependent scaling solution, and the $1/\bar{k}^{2}$ factor ensures correct normalisation.

\begin{figure}
\begin{center}
\includegraphics[width=0.49\textwidth]{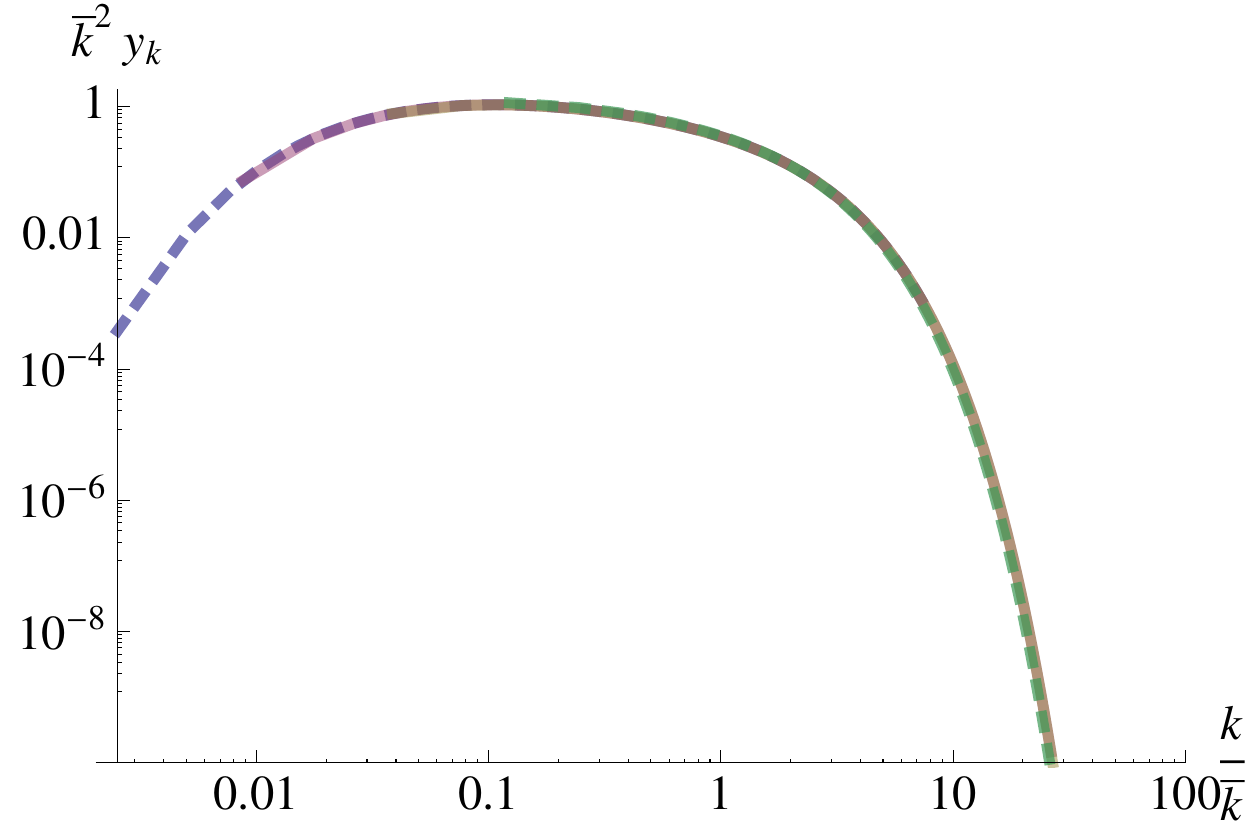}
\includegraphics[width=0.49\textwidth]{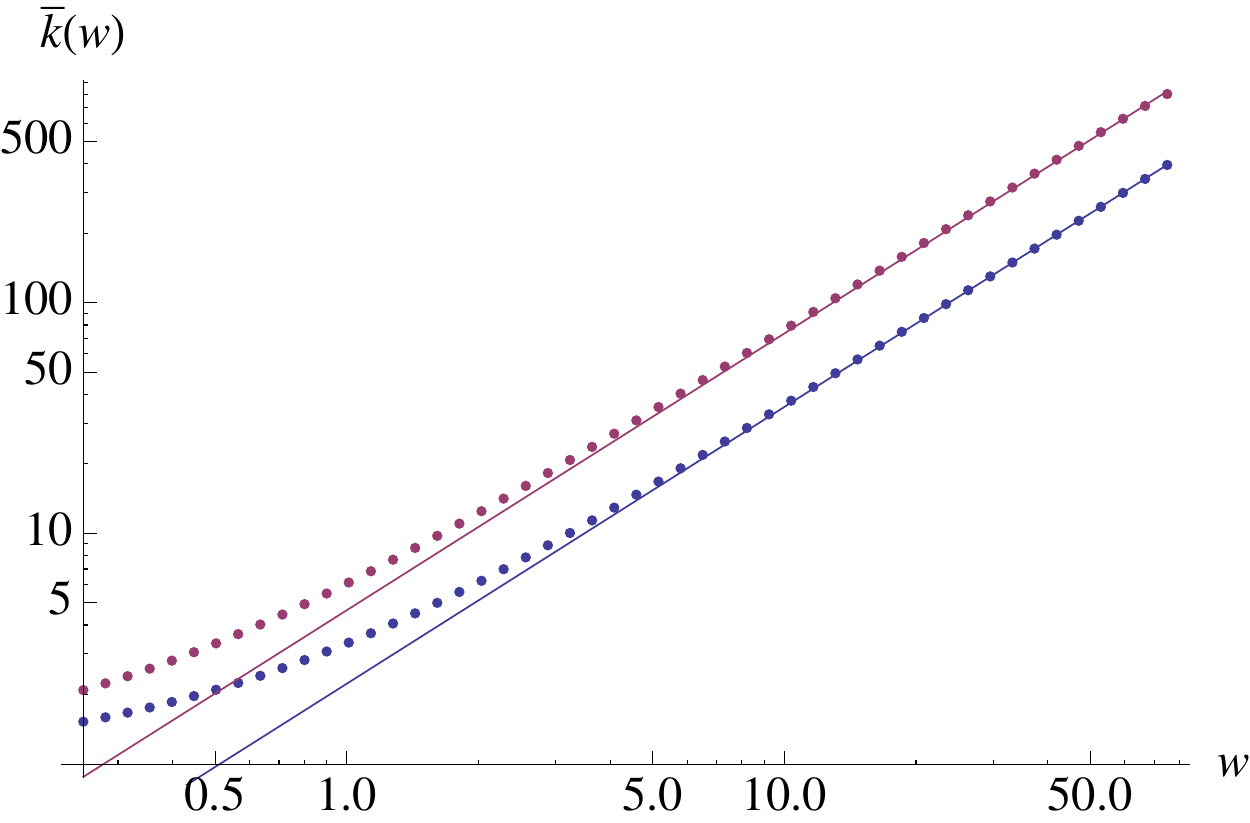}
\caption{Illustration of convergence of aggregation equation solutions
to `scaling' behaviour. 
\emph{Left}: Solutions of aggregation equation after different
`times' $w$ (with $y_1(0) = 1$), plotted on scaled axes $\bar{k}^2 y_k$ versus
$k/\bar{k}$, where $\bar{k} = 1/\sum_k y_k$.
Green dashed line shows solution at $w = 3$, yellow at $w = 8$, red at $w = 25$,
and blue dashed at $w = 75$. 
The almost complete overlap shows that convergence to the attractor
form of the `scaling solution' is rapid.
\emph{Right}: Plot of $\bar{k}$ behaviour with time---dots
are numerical $\bar{k}(w)$ values, lines are $w^{6/5}$ curves.
Blue are for $\bar{k}(w) \equiv 1/\sum_k y_k$, red are for $\bar{k}(w) \equiv \sum_k k^2 y_k$.
}
\label{fig:sdist}
\end{center}
\end{figure}

In our case, we can check whether the numerical solutions in
Figure~\ref{fig:ndist} obey this scaling behaviour by choosing
a plausible form for $\bar{k}$ (e.g.\ for a peaked number distribution,
we could take one over the total number of DN, $(\sum_k y_k)^{-1}$),
and plotting 
$\bar{k}^2 y_k$ against $k/\bar{k}$, as shown
in Figure~\ref{fig:sdist}. We see that, for $y_1(0) = 1$ initial
conditions, the distribution converges very quickly to such a scaling solution, which
can then be used to extrapolate to larger $w$ values.
From the mathematical theory of aggregation developed in other contexts~\cite{Ernst1988,Leyvraz2006}, we expect
$f(x)$ to drop as $\sim x^{-\lambda} e^{-{\rm const.}\ x}$ for $x \gg 1$, and to have the form $f(x) \sim
x^\theta e^{-{\rm const.}/\sqrt{x}}$ for $x \ll 1$, with $\theta$ some constant, both of which match
sensibly onto the numerically obtained form.  Also, the behaviour of
$\bar{k}(w)$ with $w$ matches, for larger $w$, the $\propto w^{6/5}$ form
predicted from $\lambda = 1/6$. This also holds for different choices of $\bar{k}$, as expected for
a peaked number distribution.

\subsection{(In-)Dependence on initial conditions}
\label{sec:initialcond}

As discussed in Section~\ref{sec:dn}, we do not expect to
start the fusions-only aggregation process with single-nucleon-only
initial conditions---instead, we will have whatever was produced
during the phase when dissociations were still important.
Furthermore, fusion cross sections between small DN
will probably not be well approximated by the geometrical
scaling appropriate to large DN; for example, SM cross
sections between small nuclei display very complicated behaviour.
Since, as we discuss in the next section, there is only a finite
$w$ time available for aggregation due to Hubble expansion, the question is whether
these initial conditions, and small-$k$ behaviour, converge to the scaling solution
quickly enough for appropriate measures of convergence.

\begin{figure}
\begin{center}
\includegraphics[width=0.49\textwidth]{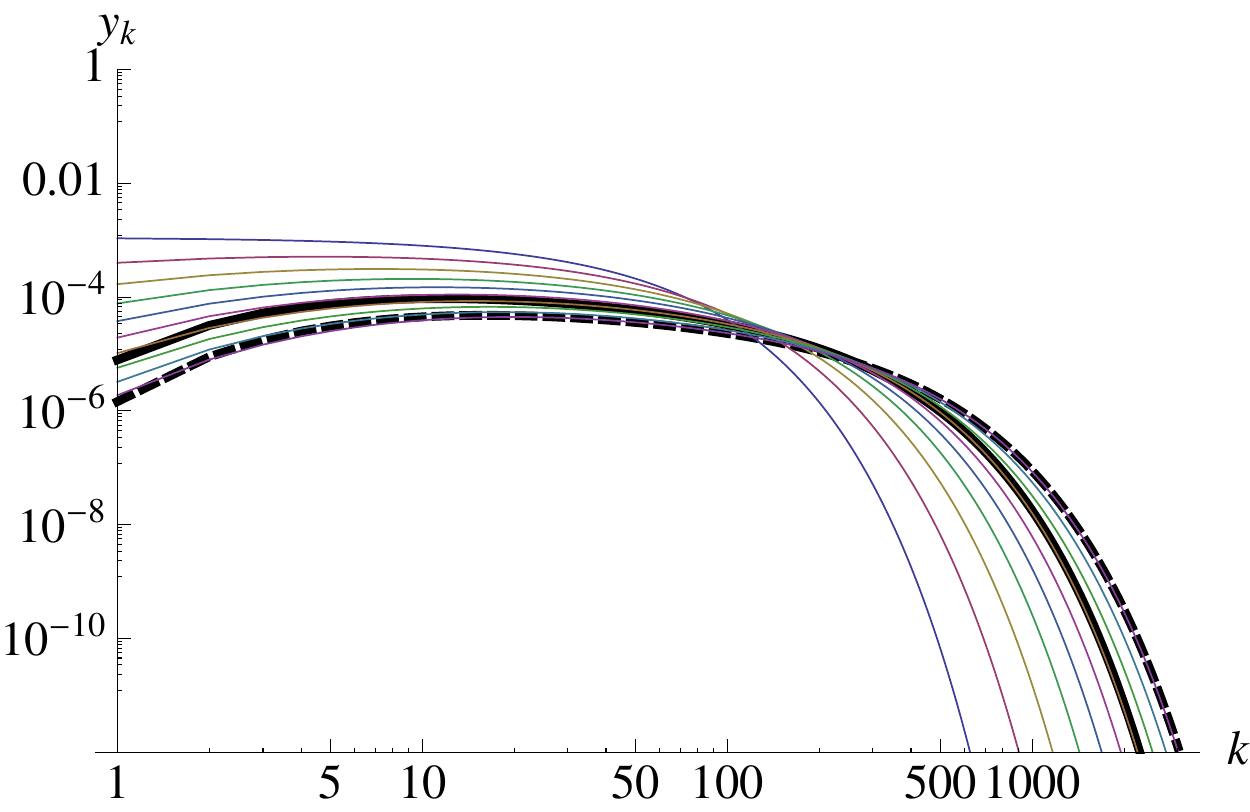}
\includegraphics[width=0.49\textwidth]{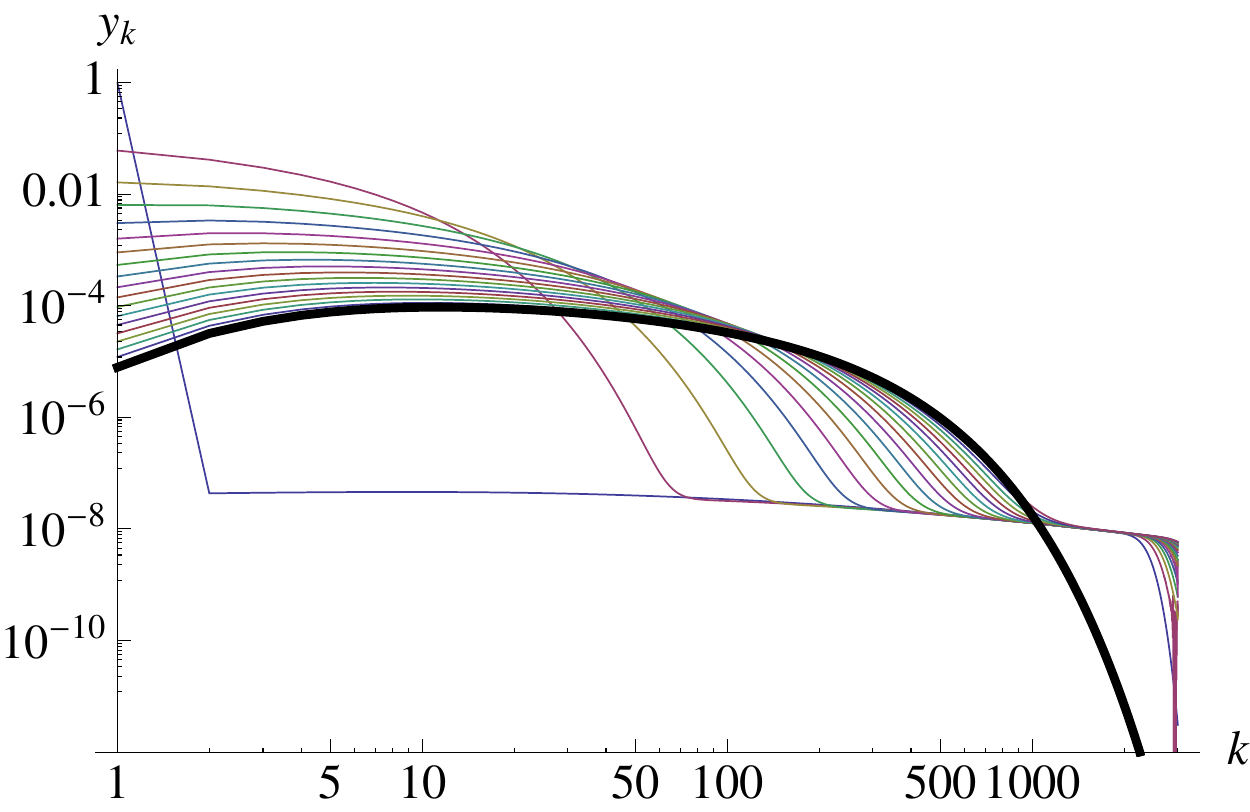}
\caption{Illustration of convergence of aggregation equation solutions
to `scaling' behaviour, for more complicated initial conditions.
\emph{Left}: Coloured curves are solutions of aggregation
dynamics, starting from $y_k(0) = e^{-k/30}$ initial conditions, at
equally-spaced $w$ values from 0 to 25.  For comparison solid black curve shows
solution at $w=25$ for monomer-only initial conditions,
while black dashed curve is monomer-only solution at the later time $w=33.25$.
We see that despite the change in initial conditions we still end up
with the scaling distribution but now it is slightly shifted in $w$-evolution
compared to the monomer-only case.
\emph{Right}: Solutions of aggregation dynamics at
equally-spaced $w$ values from 0 to 25,
starting from initial conditions with most of the mass in single nucleons
($y_1(0) = 0.97$), but also a sub-dominant population
in a broad tail out to large $k$ values (as might arise
from e.g.\ build-up while most of the nucleons are trapped
behind a low-$k$ bottleneck).
The black curve is the solution at $w = 25$ for monomer-only initial conditions.
Since the initial state has only a small fraction of the nucleons in
the tail, and fusions between states in the tail are frozen out,
the tail has no significant effect on the majority of the scaling solution.
}
\label{fig:ictest}
\end{center}
\end{figure}

Generally, initial conditions can be separated into a component for
which similar-size fusions are not frozen out, and one for which they are
(this component should be sub-dominant, otherwise aggregation does not significantly affect
the distribution at all).
Treating the latter case, in which the initial conditions include a subdominant large-$k$ `tail' for
which large~$+$~large fusions are frozen out (see the right panel
plot of Figure~\ref{fig:ictest} as an example), then as long as the aggregation
of smaller DN proceeds not much slower than the scaling behaviour,
only a small proportion of the small DN fuse with those in the tail.
To see this we can approximate the tail by purely $A$-DN, and use the fact that the
rate at which a given small $k$-DN fuses into the tail is $\Gamma_{k+A} \sim \delta \sigma_1 v_1 n_A A^{2/3} k^{-1/2}$.
So, writing the fraction of the DN in the tail as $\alpha = A n_A/n_0$,
\begin{equation}
\frac{d (1-\alpha)}{dt} \sim \Gamma_{k+a} (1-\alpha)
\quad \Rightarrow \quad
\frac{d\log(1-\alpha)}{dw} \sim - \delta \frac{n_A}{n_0} A^{2/3} k^{-1/2} ~.
\end{equation}
While $\alpha$ is small, so the small-$k$ build-up proceeds
like the scaling solution, we have $k \sim w^{6/5}$.
Since $A+A$ fusions are frozen out, $n_A$ is roughly constant throughout, so
\begin{equation}
\frac{d\alpha}{dw} \sim \delta \left(\frac{n_A}{n_0}\right)^{1/3} \alpha^{2/3} w^{-3/5}
\quad \Rightarrow \quad
\alpha_{\max} \sim \delta \frac{n_A}{n_0} w_{\max}^{6/5} \sim \delta \frac{n_A}{n_0} k_{\max} \,.
\end{equation}
Thus, either $\delta k_{\max} \gtrsim A$, in which case the tail is subsumed into
the scaling distribution, or else $\alpha$ is always $\ll 1$.
Note that this only occurs because $k$ increases sufficiently quickly with $w$.
As we shall see in Section~\ref{sec:bn}, if that does not happen,
then none of the DN may ever reach the scaling regime,
and $\alpha_{max} \sim \frac{n_A}{n_0} w_{\max}^3$ can become of order 1,
so all of the small DN can be absorbed by the frozen-out tail.

Dealing with the other case, suppose that the initial conditions have some component for
which similar-size fusions are not frozen out, e.g.\ the
left panel of Figure~\ref{fig:ictest}. If these have the same
average size as a monomer-only scaling solution after $\delta w$, then
the eventual distribution will be close to the monomer-only solution
after $w_{\max} + \delta w$. Since small $+$ large fusions are faster
than large $+$ large, the `memory' of the initial shape is erased fairly
efficiently.\footnote{Similarly, if the small-$k$ cross sections are larger than
those extrapolated down from large $k$, the effect is to start the
large-$k$ process slightly earlier, $w \rightarrow w + \delta w$, while if the
cross sections are smaller than the extrapolation then the solution
interpolates between the scaling and bottlenecked regimes---see
Section~\ref{sec:bn}.}

Figure~\ref{fig:ictest} illustrates these behaviours numerically,
with the left panel showing the convergence of non-frozen-out
initial conditions to a slightly-further-along scaling distribution,
and the right panel showing that a sub-dominant frozen-out tail
has little effect on the scaling solution obtained.

\subsection{Real-time behaviour}

The scaling distribution gives us the form of the $y_k(w)$ solution---to 
obtain the real-time distribution $y_k(t)$, we need to know the
behaviour of $w(t)$ from solving eq.(\ref{eq:wt}). Generally, we are
most interested in the $t \rightarrow \infty$ limit, so want to find
$w(t\rightarrow \infty)$. To obtain this we in turn need to know the
behaviour of $v_1(t)$, which is simple in the case that the DN are in
\emph{kinetic} equilibrium throughout the aggregation process.

The simplest way for the DN to be in kinetic equilibrium
is for them to be in thermal contact with a bath of lighter particles
during the aggregation process. If this bath is relativistic,
so $T_b \propto 1/a$ (here ignoring possible changes in the number of species for simplicity),
and taking $v_1 = \sqrt{T_b/m_1}$, we obtain, assuming radiation domination,
\begin{equation}
\frac{dw}{dT} \simeq - \left. \frac{n_0 \sigma_1 v_1}{H}\right|_{T_0} 
\left(\frac{T}{T_0}\right)^{1/2} \frac{1}{T_0} \,,
\end{equation}
where $T_0$ is the temperature when $w = 0$, i.e.\ at the start of the aggregation
process.
Solving this,
\begin{equation}
w(T) \simeq \frac{2}{3} \left. \frac{n_0 \sigma_1 v_1}{H}\right|_{T_0} \left(1 - \left(\frac{T}{T_0}\right)^{3/2}\right) \rightarrow 
\frac{2}{3}\left. \frac{n_0 \sigma_1 v_1}{H}\right|_{T_0} \,,
\label{eq:wTrel}
\end{equation}
where the limit is $T \searrow 0$.
The right panel of Figure~\ref{fig:ndist} shows the solutions
obtained at different $t$ values, assuming this relationship between
$w$ and $T$, illustrating the convergence towards the $w = w_{\max}$
solution. 

Alternatively, if the hidden sector bath is non-relativistic, so $T_b \propto a^{-2}$, we obtain
\begin{equation}
w(T) \simeq \frac{1}{2}\left. \frac{n_0 \sigma_1 v_1}{H}\right|_{T_0} \left(1 - \left(\frac{T}{T_0}\right)^{2}\right) \rightarrow \frac{1}{2}\left. \frac{n_0 \sigma_1 v_1}{H}\right|_{T_0} \,.
\label{eq:wTnr}
\end{equation}
Since $\bar{k} \sim w^{6/5}$, we have $\bar{k}(t \rightarrow \infty) \sim
\left(\left. \frac{n_0 \sigma_1 v_1}{H}\right|_{T_0}\right)^{6/5}$ in both cases.
This agrees with the approximate freeze-out calculation of
eq.(\ref{eq:foscaling}), which had $\Gamma/H \sim \frac{n_0 \sigma_1 v_1}{H} A^{-5/6}$.

Generally, since from eq.(\ref{eq:wt}) $\frac{dw}{d\log
a} = \frac{n_0 \sigma_1 v_1}{H}$, and during the radiation
era we have $H \propto a^{-2}$, $n_0 \propto a^{-3}$, and
writing $v_1 \propto a^{-\gamma}$, we have
$\frac{dw}{d\log a} \sim a^{-(1 + \gamma)}$, so,
$\Delta w \sim \Delta \left( a^{-(1 + \gamma)}\right)$
as we obtained in eqs.(\ref{eq:wTrel}) and~(\ref{eq:wTnr}).
Thus the {\it bulk of the aggregation process takes of order a Hubble time to
complete}. This is illustrated in the right panel of Figure~\ref{fig:ndist},
which shows solutions at half-$e$-folding-time intervals.
Such behaviour just comes from the freeze-out properties
of the interactions, so in the bottlenecked regime we take at
most this long as well---in fact, as we shall see in the next section,
that process may take much less than a Hubble time.


\subsection{Bottlenecked regime}
\label{sec:bn}

\begin{figure}
\begin{center}
\includegraphics[width=0.49\textwidth]{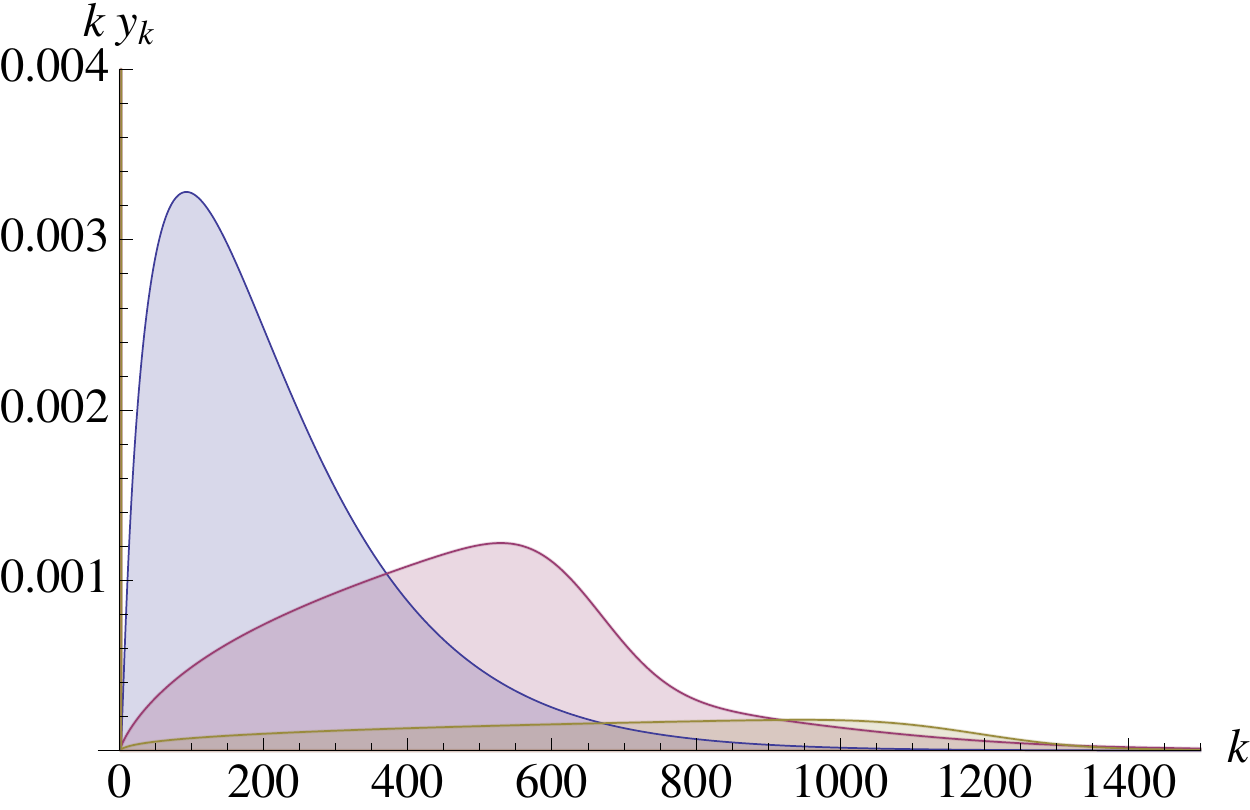}
\includegraphics[width=0.49\textwidth]{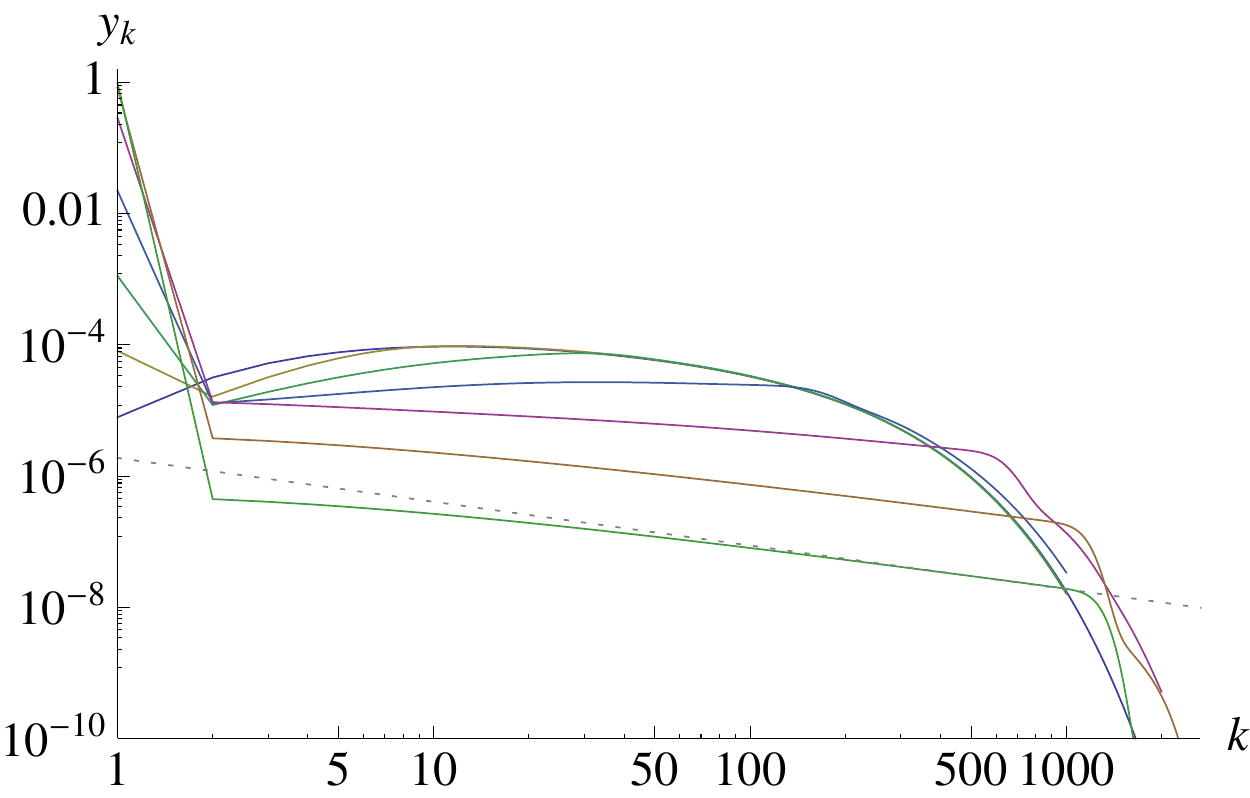}
\caption{Illustration of transition between scaling regime and bottlenecked regime.
\emph{Left:} Mass distribution at $w = 25$, starting from initial conditions
of single nucleons, $y_k(0) = \delta_{k1}$, for the original kernel with $K_{1,1} = 4$ (purple),
and modified kernels with $K_{1,1} = 10^{-4}$ (red), and $K_{1,1} = 10^{-5}$ (yellow).
We see that, within this range, making $1+1$ fusions slower results in building up larger DN.
\emph{Right:} Number distribution at $w=25$, again starting from initial conditions of single nucleons,
for $K_{1,1} = (4, 1, 10^{-1}, 10^{-2}, \dots, 10^{-6})$. The dotted line aligned with
the $K_{1,1} = 10^{-6}$ (green) curve is $\propto k^{-2/3}$. 
We transition from converging very quickly to the scaling solution,
to ending up with a power-law distribution cutting off at larger $k$.
Between $K_{1,1} = 10^{-5}$ and $10^{-6}$, the maximum $k$
reached hardly increases, with the overall number density in the power-law tail
just going down---we have reached the freeze-out limit for building up large DN.
}
\label{fig:scalingbn}
\end{center}
\end{figure}

As illustrated in Section~\ref{sec:scaling}, if the fusion
rates, as parameterised by $w_{\max}$ and the $K_{i,j}$,
are not too small compared to the support of our initial conditions, then we reach the scaling
regime of the attractor solution.  However, if some of the fusion rates are reduced far enough
to `trap' a proportion of the DN in a small-$k$ region
for long enough, then we will not reach the scaling regime.
Counter-intuitively, this can result in building
up \emph{larger} DN than would otherwise have been
the case. As roughly described by eq.(\ref{eq:fobneck}), this
occurs because small $+$ large fusions are less velocity-suppressed
in kinetic equilibrium than large $+$ large fusions, so,
if there is a bath of small DN present throughout the aggregation
process, build-up interactions may freeze out at higher $A$.
Figure~\ref{fig:scalingbn} shows a simple example
of moving between the scaling and bottlenecked regimes,
which may be helpful to keep in mind through the following.

Since small $+$ large fusions keep the number density
of large DN the same, they act to increase the rate of large $+$ large fusions,
which goes as $A^{2/3} A^{-1/2} = A^{1/6}$.
If $\Gamma/H$ becomes of order 1 or higher, then large-large
fusions will start operating and bring it down to $\sim 1$, establishing
a scaling distribution for the larger DN. Since the rate of fusions for 
a single small DN is larger than that for large DN, this also
means that all of the small DN will be used up, clearing the bottleneck
and placing us in the scaling regime.

The more interesting case is when large-large interactions are always
frozen out. We can then model the aggregation process as a combination
of the slow creation of large post-bottleneck `seed' DN, and the fast
accretion of small pre-bottleneck DN onto these. Looking first
at the accretion process, from our previous assumptions the growth
rate for a large DN scales geometrically, $\propto R^2 \sim A^{2/3}$.
So, $dk/dw \sim k^{2/3} y_s$, where $y_s$ parameterises the concentration
(and size)
of small DN, and thus $k \sim \left( \int dw \, y_s \right)^3$.
If the bath of small DN is populated throughout most of the $w$ time, then
$k_{\max} \sim w_{\max}^3$, realising the freeze-out bound of eq.(\ref{eq:fobneck}).

As well as the maximum size, we are also interested in the number
distribution over $k$. If a seed DN is produced at time $w_{\rm inj}$,
then its eventual size will (for roughly constant $y_s$) be $k \sim
(w_{\max} - w_{\rm inj})^3$. More precisely, if we change time variable
to $z$, where $dz/dw = y_s$, we have $k \sim (z_{\max} - z_{\rm
inj})^3$. This means that the relationship between injection time
$z_{\rm inj}$ and final size $k$ is given by $-\frac{dz_{\rm inj}}{dk} \sim k^{-2/3}$.
Next, write $a_k(z)$ for the total concentration of seed $k$-DN injected by
time $z$. Since accretion only changes the $k$-number, and not the density,
of an injected seed population, we obtain
\begin{equation}
y_k = -\frac{d}{dk}\left(\sum_j a_j(z_{\rm inj}(j \rightarrow k))\right)
=
-\sum_j \frac{d a_j}{dz} \frac{dz_{\rm inj}(j \rightarrow k)}{dk}
\sim k^{-2/3} \sum_j c_j \frac{d a_j}{dz}
\end{equation}
For $k$ large compared to the $j$ which dominate the sum,
we can write this as $k^{-2/3} f_i(z_{\rm inj})$.
In particular, if for small $z$ the injection rate is roughly
constant (this is reasonable, since only a small fraction of the pre-bottleneck bath
has been used up, and the temperature, scale factor etc. change by only
a small amount), then for large $k$ we should have a power law
number distribution $y_k \sim k^{-2/3}$.

As we increase the injection rate $f_i$, we move through the
three different regimes described at the end of Section~\ref{sec:fusionfo}:
\begin{itemize}
\item  If the injection rate is small enough, then most of the small
bath is not used, and most of the mass is stuck behind the bottleneck,
with a small proportion in a $y_k \sim k^{-2/3} f_i(z_{\rm inj})$ tail,
which extends to $k_{\max} \sim w_{\max}^3$. As an example, if we assume that
$f_i$ decreases with time, such that the nucleon number integral
$\int^{k_{\max}} dk\, k \, k^{-2/3} f_i$ is dominated by the upper
decade, $\sim k_{\max}^{4/3} f_i(0)$, then the upper limit of this regime
is given by $f_i(0) \sim k_{\max}^{-4/3} \sim w_{\max}^{-4}$.
\item For larger injection rates, we use up all of the small bath before
we reach $w_{\max}$, so we build up to correspondingly smaller $k_{\max}$.
Under the same assumptions on $f_i$, we have $k_{\max} \sim f_i(0)^{-3/4}$.
\item For sufficiently large injection rates, large-large fusions eventually become
important, and we enter the scaling regime. We expect this to happen roughly when
$k_{\max}$ for the addition process is comparable to the $k$ values of the scaling
peak. Since this has $\bar{k} \sim w^{6/5}$, under the above assumptions
on $f_i$ we expect the cross-over to be at around $f_i(0) \sim w_{\max}^{-8/5}$.
\end{itemize}

In the first regime, the proportion of DM in the pre-bottleneck bath stays
roughly constant, so $\frac{dz}{dw} = y_s \simeq {\rm const}$.
From the previous section, $\Delta w \sim \Delta (a^{-(1 + \gamma)})$,
so the bulk of the process again takes of order a Hubble time.
For the intermediate regime, we use up most of the small bath within
$\Delta w \simeq (3 k_{\rm max})^{1/3} < w_{\rm max}$,
a small fraction of a Hubble time. For the rest of $w$ time,
either large-large fusions are frozen out, in which case we only 
make small modifications to the number distribution,
or we enter the scaling regime.

Figure~\ref{fig:scalingbn} shows the numerical solution of a particularly
simple bottlenecked example---using the geometrical kernel of 
eq.(\ref{eq:ker1}), but reducing $K_{1,1}$. In this case,
the injection rate $f_i \propto y_1^2$, so for most of the addition
process it is constant, giving a power-law number distribution $y_k \sim k^{-2/3}$
for large $k$. The right panel of Figure~\ref{fig:scalingbn} illustrates how we move through
the three regimes identified above as we decrease $K_{1,1}$. We start
out converging quickly to the scaling distribution, but for
$K_{1,1} \lesssim 10^{-4}$ we never reach this regime, ending up with
a power law out to larger $k$. For $K_{1,1} \lesssim 10^{-5}$,
most of the $k=1$ population never makes it past the bottleneck,
and our power-law tail goes to smaller concentrations
rather than higher $k$ values.

\begin{figure}
\begin{center}
\includegraphics[width=0.49\textwidth]{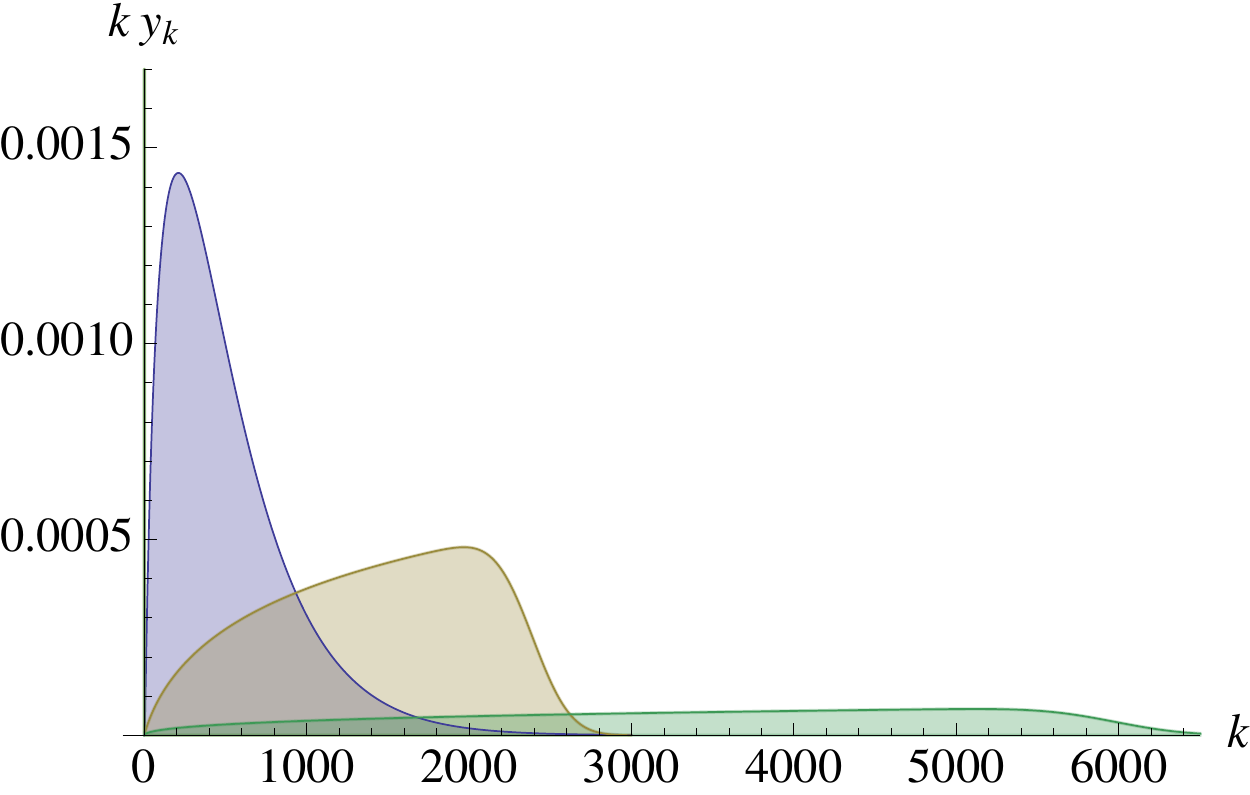}
\includegraphics[width=0.49\textwidth]{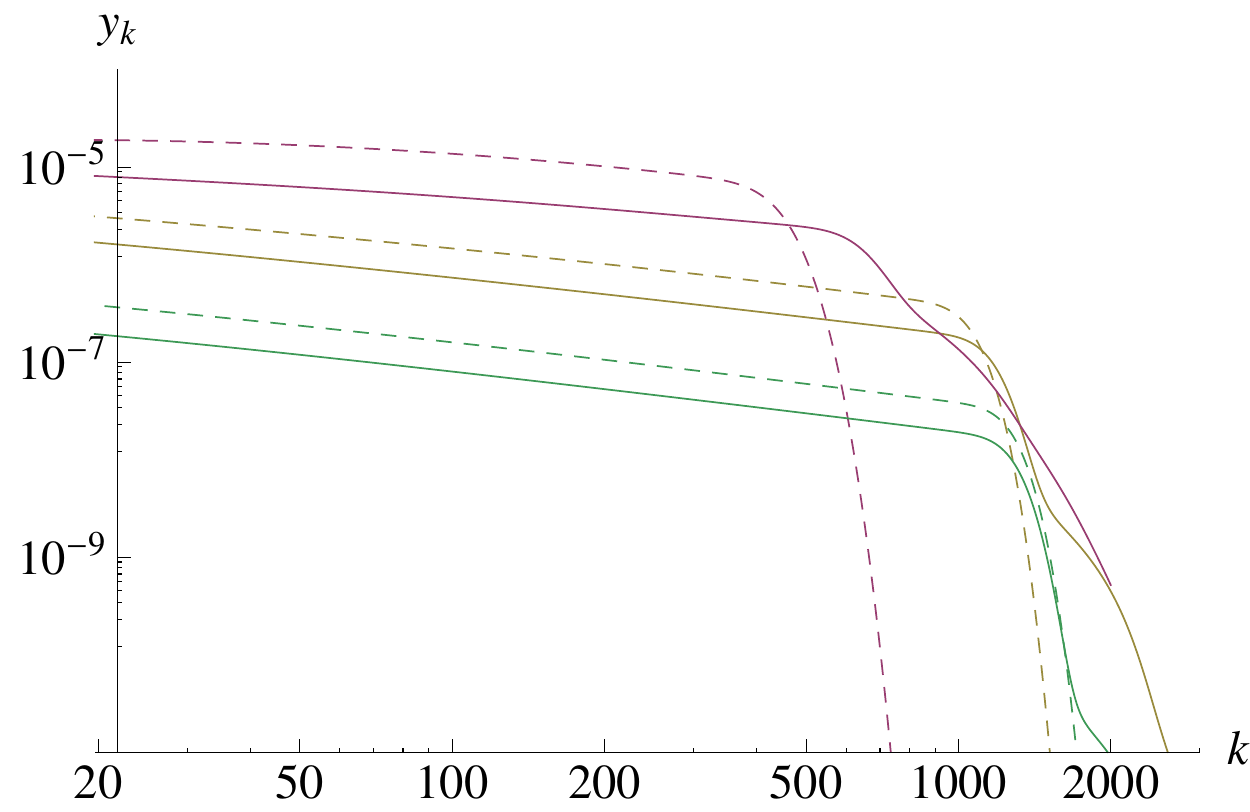}
\caption{Further illustrations of behaviour in the bottlenecked regime.
\emph{Left:} Mass distribution at $w = 50$, starting
from single nucleons, for original kernel $K_{1,1} = 4$ (purple),
and modified kernels with
$K_{1,1} = 10^{-5}$ (yellow), $K_{1,1} = 10^{-6}$ (green) (latter two
in the `addition approximation', i.e. only taking into account
$1 + k \rightarrow (1+k)$ fusions). Comparing to
Figure~\ref{fig:scalingbn}, going to larger $w$ increases the difference
between the bottlenecked and scaling solutions.
\emph{Right:} Solid lines are number distributions
at $w = 25$, starting from single nucleons, for $K_{1,1} = 10^{-4}$ (red), $K_{1,1} = 10^{-5}$ (yellow), and 
$K_{1,1} = 10^{-6}$ (green). Dashed lines are number distributions
in the addition approximation.
}
\label{fig:1plus}
\end{center}
\end{figure} 

The right panel of Figure~\ref{fig:1plus} shows explicitly that,
for the bottlenecked solutions, we are close to being in the
addition-dominated regime, i.e.\ the dominant process is $1 + k \rightarrow (k+1)$.
Comparing the left panel to Figure~\ref{fig:scalingbn} illustrates that, as we increase $w_{\max}$,
the difference between the $k_{\max}$ attainable in the
scaling and bottlenecked regimes increases.

As was the case in the scaling regime, sufficiently small changes in the
initial conditions, or in the rates of individual processes, will not
make a major difference to the eventual number distribution. If we start
with a small mass fraction of the DN past the bottleneck, this will be
equivalent to an injection spike at $z=0$, and so a bump at the end
of the overall tail. If some of the rates for small $+$ large fusions
differ from the geometrical approximation, say for large DN of size
$k$, then this will affect all of the large DN that grow to sizes $> k$---as 
long as the reduction in the rate is not large enough to cause a
further bottleneck, there will be little change in the qualitative form
of the final distribution.


\section{Aspects of dark sector phenomenology}

So far, we have investigated the BBDN process,
and the number distribution of DN that may result from it.
The DM self-interactions, and the extended
hidden sector needed to realise such models, mean that these
theories generically have the possibility of interesting hidden sector
phenomenology, including non-standard indirect detection signals,
modifications of halo properties, and early-universe signatures of extra species. 
Also, though not required in these models, it is possible
that the DN have sufficiently strong interactions with 
SM states to give signals in future direct detection experiments.
If that is the case, then such signals may differ from those
of usual WIMP DM, and have different relationships
to collider bounds \cite{NDMdirect}. Additionally, such interactions
may lead to the capture of DN in astrophysical objects,
and then self-interactions among captured DN could
become important.

In standard ADM models, annihilations with the relic
symmetric population can lead to astrophysical energy injections,
most detectably in the early universe,
putting a limit on how large this population can be. In nuclear DM models
there is additionally the energy injected by inelastic processes, particularly
the release of binding energy from fusions. 

For simple ADM models with DM mass $\lesssim 100 \GeV$, annihilation of the symmetric
component generally needs to go into, or via, lighter hidden sector particles rather
than directly to the SM, to avoid constraints from
direct detection and collider experiments~\cite{MarchRussell2012}.
We will see below that in the case of DN,
the constraints from direct detection on the SM
couplings of individual DN constituents are often even tighter,
confirming the need to annihilate into lighter hidden sector particles.
In addition, if we want to build up DN via fusions from single nucleons,
we need some lighter state for at least the first few fusions to de-excite into, as
the limits on SM interactions are generally strong enough that
we cannot de-excite fast enough via those alone.
We also need some mediator particle to transmit the binding force.
If any of these additional states are sufficiently light, strongly-interacting,
or long-lived, then there may be astrophysical constraints
on their properties.

The amplitudes for sufficiently weak interactions between 
DN and other states will, for interactions involving individual dark
nucleons, factorise into a form factor and a single-nucleon amplitude.
For momentum exchanges which are small compared to the
inverse radius of the DN, the form factor will
be $\propto A$, i.e.\ the interaction will be
coherence enhanced, and will fall off in some nuclear-structure-dependent
way for larger momenta \cite{NDMdirect} analogously to the scattering
of SM nuclei with WIMP DM.

Assuming that DN scatterings with SM nuclei in 
DM direct experiments are in the small momentum transfer
regime, the coherence factor has the effect of
\emph{enhancing}, for a given SM-dark nucleon interaction
strength, the constraints from these experiments.
Taking the DN to be of a single size $A$, the number
density is $\propto 1/A$, but the scattering rate is coherence-enhanced
by $A^2$, so the overall event rate scales as $A$.
For the distributions discussed in the previous sections, both
the mass distribution $k y_k$ and the scattering-rate
distribution $k^2 y_k$ are peaked over a small logarithmic range
in $k$, so the $\propto A$ enhancement is a reasonable approximation. 
The same is true, for simple injection profiles, in the bottlenecked case.  
Since the rate for processes with only SM particles in the initial 
state is generally set by the interaction strength
with individual dark nucleons, collider constraints, 
and those from the annihilation of the symmetric
DM component in the early universe, are relatively less stringent
compared to direct detection. 

In addition to these effects on overall rates, for intermediate momentum
transfers the effect of the momentum-dependent DN form factor may be
important. The possibility of `dark form factors' influencing the shape
of direct detection recoil spectrum has been investigated by a number of
authors, often in the context of suppressing low-momentum scatterings~\cite{Chang2009,Feldstein2009},
but also addressing soft form factors~\cite{Gelmini2002,McDermott:2011hx,Cherry2014}.
Also, for sufficiently large energy transfers, inelastic processes
involving excited DN states or fissions may be important.
In a forthcoming paper \cite{NDMdirect} we investigate the possible direct
detection phenomenology of DN in quantitative detail, as well as exploring
the possible consequences of its capture in astrophysical objects.


\subsection{Post-nucleosynthesis energetics in the dark sector}
\label{sec:selfint}

Since we have been considering DM with strong
self-interactions, an obvious question is whether these
self-interactions have late-time consequences.
For elastically-scattering DM with velocity-independent
self-scattering cross section, the observational
limit on the self-scattering cross section is
$\sigma_{XX}/m_X \lesssim 1 \barn / \GeV$ (see e.g.~\cite{Tulin2013}).
For a population of $A$-DN,
\begin{equation}
\frac{\sigma_{AA}}{m_A} \simeq \frac{0.05 \barn}{\GeV} 
A^{-1/3}
\left(\frac{1 \GeV}{m_1}\right)^{1/3} 
\left(\frac{1 \GeV \fermi^{-3}} {\rho_b}\right)^{2/3} ,
\end{equation}
in the notation of eq.(\ref{eq:fon1}). From the results of
Section~\ref{sec:fusionfo}, if we scale
$m_1 \mapsto \lambda m_1$, $\rho_b \mapsto \lambda^4 \rho_b$ etc.,
then in the scaling regime, $A \mapsto \lambda^{-12/5} A$.
Thus, $\frac{\sigma_{AA}}{m_A} \mapsto \lambda^{-11/5}\frac{\sigma_{AA}}{m_A}$,
so we would hit the observational limit at $\sim 10 \MeV$ scales
for the constituents, assuming that we built up to the largest
allowed sizes.

Realistically,  the situation can be more complicated than this.
Firstly, there would be a spectrum of DNN values. However,
for the scaling distribution, and simple forms of the bottlenecked
distributions, the area distribution $k^{2/3} n_k$ and the mass
distribution $k n_k$ both receive most of their contribution
from a single decade or so of nucleon numbers,
so the above calculations will hold approximately.

More interestingly, the fact that DN collisions may not be elastic
could have phenomenological consequences. Since, neglecting collisions,
DN of different sizes will have the same velocity distribution,
with e.g. $v \sim 10^{-3}$ in the Milky Way, large DN will have kinetic energies
much larger than the single-nucleon binding energy scale. Thus, it is possible
that DN-DN collisions will result in inelastic collisions,
including nuclear fragmentations, transitions to excited states,
and fusions.

Some types of collisions may be dissipative, reducing
the average kinetic energy (KE) per nucleon of the final state DN. 
Examples include scattering to excited states, in which 
some of the initial KE is lost into de-excitation products
(fusions with sufficiently small binding energy differences
also lose KE overall).
If these de-excitation products interact with the SM,
there may be indirect detection signals from these
processes.
Independently, the inelastic form of the scattering will 
tend to lead to the contraction of DM mass distributions,
as compared to elastic scatterings.
It may be the case that, in regions of sufficiently high DM
density, there is the possibility of run-away contraction
(in particular, if the DM is self-gravitating, then
removing KE results in the distribution contracting
and also heating up, due to the negative heat capacity).
More generally, the effects and constraints will be
different from the usual simulations of elastic DM self-interactions.

On the other hand, the presumed binding-energy-based stability of
large DN means that there may be exothermic collisions, in which
the average KE of the DN is increased (c.f. fusions in stars, in the SM).
If the velocity kick given to the DN is large compared to the velocity
dispersion in the halo, exothermic collisions will have the effect of clearing out the
central high-density region, until the number density is low enough
that collisions occur once per particle per Hubble time.
Since smaller halos tend to have smaller velocity dispersions,
small velocity kicks may only modify structure on small scales
(where there are problems with $\Lambda$-CDM predictions for structure).

These general effects of inelastic DM self-interactions
are not specific to nuclear DM, and a number of models
have been investigated in the literature. In particular,
models in which the DM transforms under some approximate
symmetry can easily realise small mass splitting,
and the effects of illustrative cases corresponding to a Yukawa self-interaction
have been considered in~\cite{Loeb2010, Schutz2014}.

Turning to the energy released in de-excitations;
considering energy injections once the aggregation process has frozen
out, the injection rate is just set by the local DM velocities and number density
(which, in regions where only a small fraction of the DN undergo 
self-interactions, will be the standard collisionless DM values).
Generally, if the rate is velocity-suppressed (as for large
DN-DN collisions), and freeze-out is before BBN, then present-day
cosmic ray (CR) constraints dominate those from earlier
times, if the injection is to sufficiently energetic SM
particles~\cite{Essig2013, Jedamzik2009}.
Velocity-independent processes with rate $\propto n^2$ 
have BBN, CMB and CR constraints within a few orders of magnitude of
each other, depending on the form of SM injections~\cite{Essig2013,
Madhavacheril2013, Jedamzik2009}.

In regions where only a small fraction of DN undergo fusions,
the proportion of the DM mass density represented
by the binding energy released in late-time fusions is 
\begin{align}
\langle \sigma v \rangle n_A t_{\rm sys} \frac{\Delta BE}{M_A}
\sim \, 
& 10^{-3} A^{-2/3}
\frac{\Delta BE}{A^{2/3} 0.01 m_1} 
\Bigg\{\frac{\rho_{\rm DM}}{0.3 \GeV \cm^{-3}}  \nonumber \\
&\left.\left(\frac{1 \GeV \fermi^{-3}}{\rho_b}\right)^{2/3}
\left(\frac{1 \GeV}{m_1}\right)^{1/3}
\left(\frac{v}{10^{-3}}\right)
\left(\frac{t_{\rm sys}}{10 \Gyr}\right) \right\} \,,
\label{eq:latebe}
\end{align}
(where the $\Delta BE$ terms corresponds to the binding energy
difference in $A + A \rightarrow 2 A$ fusions scaling as $A^{2/3}
\beta$, in the notation of eq.(\ref{eq:semf}); $\beta = 0.01
m_1$ is around the SM value).
This corresponds to a proportion $\sim 0.1 A^{-1/3} \{\ \}$
of the DN undergoing collisions, where $\{\ \}$ indicates
the term in curly brackets in equation~\ref{eq:latebe}, so the total KE of the DN
involved is $\sim 7 \times 10^{-8} A^{-1/3} \{\ \}$ of the 
DM mass density.

As an illustration of a comparable indirect detection signal, for
standard symmetric DM, the proportion of the DM mass density released in
$s$-wave annihilations is
\begin{equation}
\langle \sigma v\rangle n_X t_{\rm gal} \sim 3 \times 10^{-8} \left(\frac{100 \MeV}{m_X}\right)
\left(\frac{\langle \sigma v\rangle_X}{\pb}\right)\left(
\frac{\rho_{\rm DM}}{0.3 \GeV \cm^{-3}}\right) \,,
\end{equation}
where we take $t_{\rm gal} \sim 10 \Gyr$.
Since the detectability of CR signals from annihilating DM at $\sim
\pb$ cross-sections depends on the SM injection channel etc.~\cite{Cirelli2012},
the same applies for DN collisions within the $A$ ranges of interest.
Given the range of initial DN masses, and the possibility of 
excited states of DN, such signals may have a richer structure than
the indirect detection signals usually considered.
Additionally, the geometric cross sections we have assumed give different
velocity dependence to the partial wave processes
usually considered.

Also, as considered above, for dissipative collisions in sufficiently
dense regions, there may be the possibility of (run-away) contraction
of the DN distribution, which could significantly affect signals from e.g.
the Galactic centre.


\subsection{Light dark sector states}
\label{sec:lightdark}

As mentioned above, DN models generally require additional lighter
hidden sector states. Since the assumption is that these do not make up
the bulk of the DM, they either need to never have a large abundance
(generally difficult to realise in thermal ADM histories),
be sufficiently light and weakly interacting with the SM to persist as
dark radiation, or their abundance needs to be reduced.

Reducing the yield of a species can occur by transferring its
energy density to other hidden sector species, or 
to the SM. In the latter case, if this injection occurs during or
after BBN, there are constraints on its form and magnitude.
For injection to hadronic channels, the total energy density injected
after $T_\gamma \sim 1 \MeV$ must be significantly below that of the DM.
For electromagnetic ($e^\pm$ or $\gamma$) injection, the dominant effect at times before
thermalisation becomes inefficient, $\lesssim 10^4 \second$, is
alteration of the photon-baryon ratio,\footnote{and also increasing the photon
temperature relative to that of neutrinos, so decreasing $N_{\rm eff}$
at later times compared to BBN.} which constrains the amount of energy injected
to $\lesssim 0.1 \rho_\gamma$~\cite{Simha2008}.
At later times, any energy injections, apart from those
to neutrinos, must again be significantly
sub-DM~\cite{Jedamzik2006, Hu1993,Slatyer2012}.

For hidden sector particles of mass $\gg 100 \MeV$,
their (symmetric) chemical equilibrium abundance (at the SM temperature)
by BBN times is significantly sub-DM. The dominant
constraints on their couplings to the SM generally come
from collider experiments, and permit decay times
of $\ll 1 \sec$ (e.g.~\cite{Lees2014}). So, in these cases, if such decays
are possible, then there are generally
consistent scenarios in which any initial abundance decays to
the SM before BBN.\footnote{As noted in previous sections,
obtaining sufficiently fast annihilations (directly to the SM) to reduce
the energy density to sub-DM levels is generally difficult
for $m \ll 100 \GeV$, without involving lighter states or non-minimal flavour structures.}

On the other hand, for hidden sector particles with $m \ll 100 \MeV$,
astrophysical, collider and other observations place strong constraints on their
SM couplings. These restrict their SM decay times to
$\gtrsim 1 \sec \left(\frac{100 \MeV}{m}\right)^{1 + 2k}$,
where $k \ge 0$ is set by the mass dimension of the decay operator
(e.g.~\cite{Raffelt2006, Kazanas2014}),
and mean that direct interactions with the SM are frozen out
below very high temperatures ($T_\gamma \gg m$).\footnote{Faster
decay times may be possible with more complex dark sectors,
but these generally involve additional lighter dark states.}
If $m \gtrsim 10 \keV$, purely SM decays occur
at $T_\gamma < m$, so the decay of a thermal abundance would transfer
$\gtrsim \rho_\gamma$ energy density to the SM.
For smaller $m$, the decay time is $\gtrsim 10^4 \second$,
so the constraints on energy injection to the SM are much more
severe. In either case, limits on SM energy injection generally imply
that the majority of the initial thermal energy density needs
to be transferred to other, lighter hidden sector states.\footnote{It
may be possible to realise alternative scenarios in which energy is
transferred to the SM indirectly, via other hidden sector states. If the
hidden sector is at a slightly lower temperature
than the SM, and the transfer is before $\sim 10^4 \second$,
this may be safe (sitting at the upper end of the mass range,
and transferring mostly before BBN, may also work).
This would require that number-changing interactions
with other hidden sector states are sufficiently fast
to keep the light state in thermal equilibrium during
the process. Three-body interactions with heavier
hidden sector states are suppressed by two factors
of the heavy state number density (which is small for
DM of SM mass or higher), and inelastic two-body collisions are only
relevant if such excitations are accessible at low temperatures,
imposing model-building constraints.
Alternatively, models in which SM energy injection is dominantly
to neutrinos may also be viable.} Considering these states in turn,
we eventually require that there be long-lived hidden sector states,
which will act as dark radiation, at least during the early universe.
Extra relativistic species are compatible with current observations~\cite{Ade2013}, and if the hidden sector is at a lower temperature
than the SM, its contribution to the effective number of such species
$N_{\rm eff}$ may be small.
Plausible candidates for these species within hidden sector models
include very light pseudo-Nambu-Goldstone bosons or $Z'$ states.


\section{Conclusions}
\label{sec:conclusions}

In this paper we have studied the ``Big Bang dark nucleosynthesis'' process by which
`nuclear' bound states of DM may be built up in the early universe.  Specifically
we focussed on the case of asymmetric DM models where the nuclear binding energy per
dark nucleon saturates in the large nucleon number limit.

We find that, if fusions between small dark nuclei (DN)
happen sufficiently fast, and fusion cross-sections between
large DN scale on average geometrically, the resulting number distribution
generically takes on a universal form, illustrated in Figures~\ref{fig:ndist} and~\ref{fig:sdist}.
This result is broadly independent of small 
changes to the initial conditions or fusion rates, assuming that the DN are in kinetic
equilibrium throughout the process, as discussed in Section \ref{sec:initialcond}.  The
average mass of the DN built up during this process can be as large as $\sim 10^8 \GeV$.

If fusion reactions between small DN are not large enough to
reach this regime, but large-small fusions still occur at
an appreciable rate, then there is the counter-intuitive possibility
of building up even larger DN due to the higher velocities
of smaller particles, leading to larger fusion rates.
The resulting number distribution generally takes
the form of a power law multiplied by an `injection profile'
parameterising the change of small-small fusion rates with time.
Figures~\ref{fig:scalingbn} and~\ref{fig:1plus} illustrate this
behaviour.

Given this possibility of building up large dark nuclear bound states,
direct detection signals may be modified in a number of ways---by the
coherent enhancement of SM-DM interactions, by
dark form factors if the DN radius is significantly larger than SM nuclear radii,
and by inelastic interactions if there are sufficiently
low-lying excitations. We reserve detailed discussion
of such direct detection phenomenology to a forthcoming paper~\cite{NDMdirect}.

The possibility of inelastic collisions can also lead to interesting
astrophysical dark sector interactions as mentioned in Section~\ref{sec:selfint}.
These include annihilation-type indirect detection signals,
which are usually absent from asymmetric DM models
(and here may have a richer structure, corresponding to
the number distribution of DN).  Inelastic collisions
may also modify the effects of DM self-interactions
on halo structure, especially at short distance scales.
In addition, as discussed in~\cite{NDMdirect}, there are a range 
of possible consequences for the capture of DN by astrophysical objects
as well, from ejection by de-excitations, through to fusions leading to a very dense DN core.

Finally, the discussion in this paper has intentionally been as
model-independent as is practical, investigating idealised
versions of the behaviour that classes of models can display.
The physics is inspired by that observed in the Standard Model (SM)
sector, with the simple, but important, change that the Coulomb barrier
is removed.  Although the example of the SM reassures us that a sufficiently
complicated model can realise the physics we discuss, it would be interesting to
investigate, along the lines of \cite{Hashimoto2011,Detmold2014ii,Krnjaic2014}, specific simple toy
models which realise all or some of the features discussed in this paper. As with the SM, we would
expect there to be additional features in the dark nuclear spectroscopy and interaction cross
sections, e.g.\ due to shell structure, on top of the general scaling behaviour of nuclear properties that
we have utilised.  In addition the extra hidden sector states
required (to mediate binding forces, carry away de-excitation energy, etc)
beyond the dark nucleon itself can be relevant for astrophysical phenomenology
and constraints.  In particular, if some of these states have masses $\lesssim 100 \MeV$,
there is likely to be a need for extra hidden sector states which
act as dark radiation.  Another avenue for further work is the possibility
of symmetric DM models.


\section*{Acknowledgements}

We wish to thank Asimina Arvanitaki, Masha Baryakhtar, Peter Graham, Felix Kahlhoefer, and Surjeet Rajendran
for discussions.  EH, RL and JMR thank the CERN Theory Group,
and SMW thanks the Oxford Physics Department, for hospitality
during completion of this work.  This work was supported in part by ERC grant
BSMOXFORD no. 228169.  EH and RL acknowledge support from STFC studentships, and
in addition, RL is grateful for support from the Buckee Scholarship at Merton College,
Oxford.


\appendix


\section{\label{app:chemeq}Transition from equilibrium to aggregation}

In chemical equilibrium, detailed balance for the
$k + (A - k) \leftrightarrow A$ process implies that
\begin{equation}
\Gamma_{A \rightarrow (k,A-k)} \tn_A = \langle \sigma v \rangle_{(k,A-k) \rightarrow A} 
\tn_k \tn_{A-k} \,,
\end{equation}
where the $\tn_k$ are the equilibrium number densities, $\Gamma_{A
\rightarrow (k,A-k)}$ is the rate of $A \rightarrow (k,A-k)$ dissociations
for a single $A$-DN (in general, this depends
on the abundances and velocities of other species), and $\langle \sigma v \rangle_{(k,A-k) \rightarrow A}$
is the thermally averaged fusion cross section (we
assume, as is almost always the case, that mean free paths are long 
enough so that velocity rather than diffusivity is important).
If we assume that the fission rate
is purely determined by the temperature, then when
we are in kinetic equilibrium at (non-equilibrium) number densities $n_k$ etc.,
\begin{equation}
\Gamma = \langle \sigma v \rangle \frac{\tn_k \tn_{A-k}}{\tn_A} \,.
\end{equation}
This assumption about the fission rate will be violated if DN-DN
collisions often lead to `prompt' fragmentations, rather than either
elastic scatterings, or inelastic collisions that result in a
thermally-equilibrated excited state (c.f.\ the `compound nucleus' model
for SM nuclear collisions~\cite{Bertulani2007}). The same is true for DN collisions
with other baths that do not have their chemical equilibrium abundance.

Assuming ideal gas behaviour, the equilibrium number densities
are related by 
\begin{equation}
\frac{\tn_A}{\tn_{A-k} \tn_k} = \frac{g_A}{g_k g_{A-k}}
\left( \frac{2 \pi}{m_A T} \frac{m_A^2}{m_k m_{A-k}}\right)^{3/2} 
e^{(m_k + m_{A-k} - m_A)/T} \,,
\end{equation}
where $g_k$ is the number of effective degrees of freedom of $k$-DN.
Writing $m_k = k m_1 - B_k$, and assuming that binding energies
are a small fraction of the mass, this is
\begin{equation}
\sim \frac{g_A}{g_k g_{A-k}}\left(\frac{A}{k(A-k)}\right)^{3/2} \Lambda^3 
e^{(B_A - B_k - B_{A-k})/T} \,,
\end{equation}
where $\Lambda = \sqrt{2\pi/(m_1 T)}$.

The overall forward rate for the $k + (A-k) \leftrightarrow A$ process
is 
\begin{equation}
\langle \sigma v\rangle n_k n_{A-k} - \Gamma n_A
= \langle \sigma v \rangle n_k n_{A-k} \left(
1 - \frac{\tn_k \tn_{A-k}}{\tn_A}\frac{n_A}{n_k n_{A-k}}\right) \,.
\end{equation}
Thus, as long as 
\begin{equation}
\frac{n_k n_{A-k}}{n_A} \gg\frac{\tn_k \tn_{A-k}}{\tn_A}
\quad \Leftrightarrow \quad
n_k \Lambda^3 e^{\Delta B/T} \gg \frac{n_A}{n_{A-k}}\frac{g_k g_{A-k}}{g_A}
\left(\frac{k(A-k)}{A}\right)^{3/2} ,
\end{equation}
dissociations will give only a small correction to the fusion rate.
For the $1 + 1 \leftrightarrow 2$ process, we start building up
when $n_1 \Lambda^3 e^{B_2/T} > 1$. More generally,
if we assume that typical binding energy differences rise
fast enough with size difference (for example, in a liquid-drop type
model $B_k = \alpha k - \beta k^{2/3}$, the surface tension term $\beta$
gives $B_{A} - B_{A-k} - B_k \sim \beta k^{2/3}$ for $A > k$),
then if $n_0 \Lambda^3 e^{\alpha/T} \gg 1$ (where $n_0$ is the
number density of nucleons), the inequality
should hold for all of our fusion processes between $k,A-k$ states
with reasonable density, with the possible exception
of anomalously low (or wrong-sign) binding energy differences for small-number
states.

Since $n_0 \Lambda^3 \ll 1$ (the DM is a dilute gas), we have equality
$n_0 \Lambda^3 e^{\alpha/T} = 1$ at $T \ll \alpha$. After that,
assuming that $\frac{-d \log T}{d\log a} = \mathcal{O}(1)$,
we have $\frac{d \log LHS}{d\log a} \sim \frac{\alpha}{T} \gg 1$.
So, with the possible exception of small small-$k$ $B_k$ values,
we go from being near-equilibrium, to dissociations being very suppressed,
in a time that is parametrically $H^{-1} \frac{T}{\alpha}$.

Furthermore, during this time the build-up of DN proceeds at most as
quickly as a fusions-only aggregation process, like that considered
in Section~\ref{sec:agg}. 
As explained there, in many regimes this progresses
fairly uniformly over the first Hubble time, so a modification to the
fusion rates during a small initial fraction of this time will have a
small overall effect on the result.


\bibliography{ndm}
\bibliographystyle{JHEP}

\end{document}